\documentclass[useAMS,usenatbib,usegraphicx]{mn2e}

\usepackage[normalem]{ulem}  % to use \sout for striketrough
\usepackage{soul}
\usepackage{verbatim}  % for block 'comment' environment
\usepackage{hyperref}

\def\pow#1#2{#1$\times$10$^{#2}$}

\def\micron{$\mu$m}
\def\kms{$\mathrm{km}~\mathrm{s}^{-1}$}  % km/s
\def\Kkms{$\mathrm{K~km}~\mathrm{s}^{-1}$}  % K km/s
\def\Wsqm{$\mathrm{W~m^{-2}}$}  % W/m^2

\def\pccm{$\mathrm{cm}^{-3}$}
\def\Tmb{$T_\mathrm{mb}$}  % main beam temperature
\def\Tastar{${T_A}^\star$} % "T a star"
\def\Tgas{$T_\mathrm{gas}$} % gas temperature
\def\Trot{$T_\mathrm{rot}$} % rotation temperature
\def\etamb{$\eta_\mathrm{mb}$}  % main beam efficiency
\def\etaa{$\eta_\mathrm{a}$} % aperture efficiency
\def\intintens{$\int T_\mathrm{mb} \mathrm{d}v$}
\def\Eup{$E_\mathrm{up}$}  % upper level energy
\def\Jup{$J_\mathrm{up}$}  % upper level J
\def\h{$^\mathrm{h}$}  % superscript h symbol for hour in RA coordinate
\def\m{$^\mathrm{m}$}  % superscript m symbol for minutes in RA coordinate

\def\mcol#1{\multicolumn{1}{c}{#1}} % shorthand for multicolumn environment inside tabular

\def\HH{H$_2$}

\def\twCO{$^{12}$CO}
\def\thCO{$^{13}$CO}

\def\CeiO{C$^{18}$O}

\def\water{H$_2$O}
\def\Nplus{N$^+$}

\def\CHplus{CH$^+$}
\def\Cii{\hbox{\rm [C\,{\sc ii}]}}  % [CII] forbidden lines
\def\Ci{\hbox{\rm [C\,{\sc i}]}}  % [CI] forbidden lines
\def\Oi{\hbox{\rm [O\,{\sc i}]}}  % [OI] forbidden lines
\def\Nii{\hbox{\rm [N\,{\sc ii}]}}  % [NII] forbidden lines

\usepackage{color,linegoal}  % linegoal package to break highlighted parts across lines
\definecolor{darkred}{rgb}{0.75,0.0,0.0}
\definecolor{darkblue}{rgb}{0.0,0.0,0.8}
\definecolor{orange}{rgb}{0.8,0.4,0.0}

\def\markchanges{yes}  % yes or no
\def\marked{yes}
\def\unmarked{no}
\ifx\markchanges\marked % mark changes:
	\newcommand{\removed}[1]{\textcolor{darkred}{[}\sout{#1}\textcolor{darkred}{]}}  % mark removed stuff
	
\fi
\ifx\markchanges\unmarked % DO NOT mark changes:
	\newcommand{\removed}[1]{}  % do not show removed parts at all
	
\fi

\title[Warm carbon monoxide in protoplanetary discs]{Signatures of warm carbon monoxide in protoplanetary discs observed with {\it Herschel} SPIRE\thanks{{\it Herschel} is an ESA space observatory with science instruments provided by European-led Principal Investigator consortia and with important participation from NASA.}}
\author[M.~H.~D.~van der Wiel et al.]
{M.~H.~D.~van der Wiel,$^{\! 1}$\thanks{Current address: Niels Bohr Institute \& Centre for Star and Planet Formation, University of Copenhagen, Juliane Maries Vej 30, 2100 Copenhagen {\O}. E-mail: matthijs@nbi.ku.dk  } 
D.~A.~Naylor,$^{\! 1}$
I.~Kamp,$^{\! 2}$
F.~M\'enard,$^{\! 3,4}$ % for HD100546 model (Thi+ 2011)
W.-F.~Thi,$^{\! 5,4}$ % for HD100546 model (Thi+ 2011)
\newauthor
P.~Woitke,$^{\! 6}$ %  for model of HD163296 (updated from Tilling+ 2012) 
G.~Olofsson,$^{\! 7}$  % P.I. of the GT1 program
K.~M.~Pontoppidan,$^{\! 8}$  % for contributing four targets from his OT1 program
J.~Di~Francesco,$^{\! 9,10}$
A.~M.~Glauser,$^{\! 11}$
\newauthor
J.~S.~Greaves,$^{\! 6}$
R.\,J.~Ivison$^{12,13}$ \\
$^{1}$Institute for Space Imaging Science, Department of Physics \& Astronomy, University of Lethbridge, Lethbridge, AB, T1K 3M4, Canada\\
$^{2}$Kapteyn Astronomical Institute, University of Groningen, Postbus 800, 9700~AV Groningen, The Netherlands \\
$^{3}$UMI-FCA, CNRS-INSU, France (UMI 3386) and Universidad de Chile, Cerro Cal\'an, Santiago, Chile \\
$^{4}$CNRS-INSU / UJF Grenoble 1, UMR 5274, Institut de Plan\'etologie et d'Astrophysique de Grenoble (IPAG), France \\
$^{5}$Universit\'e Grenoble Alpes, Institut de Plan\'etologie et d'Astrophysique de Grenoble (IPAG), France \\
$^{6}$SUPA, School of Physics \& Astronomy, University of St~Andrews, North Haugh, St.~Andrews, KY16 9SS, UK \\
$^{7}$Department~of Astronomy, Stockholm University, AlbaNova University Center, Roslagstullsbacken 21, 10691 Stockholm, Sweden\\
$^{8}$Space Telescope Science Institute, 3700 San Martin Drive, Baltimore, MD 21218, USA\\
$^{9}$National Research Council Canada, 5071 West Saanich Road, Victoria, BC V9E 2E7, Canada\\
$^{10}$Department of Physics and Astronomy, University of Victoria, P.O.~Box 1700, STN CSC, Victoria, BC V8W 2Y2, Canada\\
$^{11}$ETH Z\"urich, Institute for Astronomy, Wolfgang-Paulistr.~27, 8093 Z\"urich, Switzerland\\
$^{12}$European Southern Observatory, Karl Schwartzchild Strasse 2,
D-85748 Garching, Germany\\
$^{13}$Institute for Astronomy, University of Edinburgh, Blackford Hill, Edinburgh EH9 3HJ, UK
\vspace{-1em}
}

\begin{document}

\date{Accepted 2014 July 20. Received 2014 July 8; in original form 2014 April 6}

\pagerange{\pageref{firstpage}--\pageref{lastpage}} \pubyear{\the\year}

\maketitle
\label{firstpage}

\begin{abstract}
% context, aims
Molecular gas constitutes the dominant mass component of protoplanetary discs. To date, these sources have not been studied comprehensively at the longest far-infrared and shortest submillimetre wavelengths.
This paper presents {\it Herschel} SPIRE~FTS spectroscopic observations towards 18 protoplanetary discs, covering the entire \mbox{450--1540~GHz} (666--195~\micron) range at $\nu/\Delta \nu \approx$ 400--1300.  
% obs. results
The spectra reveal clear detections of the dust continuum and, in six targets, a significant amount of spectral line emission primarily attributable to \twCO\ rotational lines. Other targets exhibit little to no detectable spectral lines. Low signal-to-noise detections also include signatures from \thCO, \Ci\ and HCN. For completeness, we present upper limits of non-detected lines in all targets, including low-energy transitions of \water\ and \CHplus\ molecules. 
% model comparison
The 10 \twCO\ lines that fall within the SPIRE~FTS bands trace energy levels of \mbox{$\sim$50--500~K}. Combined with lower and higher energy lines from the literature, we compare the CO rotational line energy distribution with detailed physical-chemical models, for sources where these are available and published. Our \thCO\ line detections in the disc around Herbig Be star HD~100546 exceed, by factors of $\sim$10--30, the values predicted by a model that matches a wealth of other observational constraints, including the SPIRE \twCO\ ladder. To explain the observed $^{12}$CO/\thCO\ ratio, it may be necessary to consider the combined effects of optical depth and isotope selective (photo)chemical processes. 
% 'trends' 
Considering the full sample of 18 objects, we find that the strongest line emission is observed in discs around Herbig Ae/Be stars, although not all show line emission. In addition, two of the six T~Tauri objects exhibit detectable \twCO\ lines in the SPIRE range. 
\end{abstract}

% ----- intro -------

\section{Introduction}

%\marktext  % note to referee about markup to highlight changes

% planet formation from circumstellar gas
The formation of planets and planetary systems occurs early in the evolutionary sequence of a star when the circumstellar disc is still gas-rich at stellar age \mbox{$\la$1--10~Myr} \citep[see reviews by][]{hillenbrand2008,williams2011}. Although the value of the gas-to-dust mass ratio in protoplanetary discs is uncertain \citep[e.g.,][]{meeus2010,thi2010a,bergin2013a} and may vary over time and with location in the disc, gas does represent the dominant mass component and is mainly in molecular form \citep{henning2013}. Due to the wide range of physical conditions prevailing in protoplanetary discs (e.g., total hydrogen densities $\sim$$10^4$--$10^{14}$~\pccm, temperatures $\sim$20~K to several thousand~K), observations at many different wavelengths are required to characterize fully their dust and gas content. 

% circumstellar discs around stars of various mass 
The observational proof of circumstellar discs was provided by detections of infrared excess over stellar photospheres \citep{strom1989} and by resolving kinematic signatures in CO spectral lines \citep[e.g.,][]{sargent1987}. 
While circumstellar discs have only been discovered around a few young high-mass stars ($\ga$8 $M_{\sun}$) (e.g., \citealt{kraus2010,sandell2010}; \citealt*{wang2012}; \citealt{sanchez-monge2013}), they are found to be ubiquitous around low- to intermediate-mass young stellar objects, i.e., T Tauri stars and Herbig Ae/Be stars (e.g., \citealt*{dent2005}; \citealt{silverstone2006}; \citealt{luhman2010}), in accordance with the well-known evolutionary sequence of low-mass star formation (e.g., \citealt*{shu1987}; \citealt{larson2003}).  
% differences in disc properties between HAe/Be and TT objects:
Due to lower luminosity and temperature of the central star, discs around T~Tauri stars experience lower temperatures in irradiated upper disc layers and also reach lower temperatures in their outer mid-planes (\Tgas\ down to 14~K, \citealt{akimkin2013}) than those around Herbig Ae/Be stars (\Tgas $>$ 20--25~K, \citealt{nomura2005}). In addition, polycyclic aromatic hydrocarbon (PAH) molecules are often found in discs around Herbig objects \citep[e.g.,][]{acke2010}. % and not in TT discs, Geers+ 2007; Siebenmorgen+ 2010
PAHs contribute to raising gas temperatures in irradiated gas layers through photoelectric heating. 
Consequent changes in chemistry in the gas phase and on grain surfaces may be the cause for the observed lack of diverse molecular signal in both outer and inner regions of Herbig Ae/Be discs, as compared with T~Tauri discs \citep[e.g.,][]{oberg2010c,oberg2011a,dutrey2014ppvi,pontoppidan2014ppvi}.   
Emission lines due to rotational transitions of CO, however, are routinely detected in Herbig Ae/Be objects, both from the ground in low-energy transitions \citep[e.g.,][]{dent2005,oberg2010c,guilloteau2013} and from space in high-energy transitions \citep[e.g.,][]{meeus2001,meeus2012,meeus2013}. 

Low-energy CO rotational line flux (\Jup$\leq$3), observable at wavelengths of \mbox{$\sim$0.8--3 mm} from the ground, originates primarily in the outer parts of a circumstellar gas disc. Smaller radii are probed by \Jup$\sim$15 to 30 rotational lines at wavelengths below 200~\micron\ (ISO LWS, \citealt{meeus2001}; {\it Herschel} PACS, \citealt{meeus2012,meeus2013}) and mid-infrared ro-vibrational transitions \citep[e.g.,][]{blake2004,heinbertelsen2014}. The intermediate range (3$<$\Jup$\la$15), however, has been challenging to study comprehensively, as observations are limited by narrow atmospheric transmission windows in the submillimetre or the relatively narrow instantaneous bandwidths of the space-based high spectral resolution THz spectrometer HIFI on board {\it Herschel}. 

In this paper, we present submillimetre/far-infrared measurements of 18 protoplanetary discs, 12 around Herbig Ae/Be stars and 6 around T~Tauri stars, obtained with the {\it Herschel} SPIRE (Spectral and Photometric Imaging REceiver) spectrometer ($\sim$200--700~\micron). Key parameters of the 18 sources are listed in Table~\ref{t:sourceparam}. 
The spectral lines from the CO rotational ladder presented here bridge the gap between those observable from the ground, with energy levels up to $\sim$50~K, and those above 500~K in the range of ISO \citep{kessler1996} and more recently the {\it Herschel} PACS instrument \citep{poglitsch2010}. 
Probing the full range of molecular energy levels has the advantage that radially separated origins of lines in discs allow for a physical decomposition of disc structure, even from spatially unresolved observations with single dish observatories, e.g., see \citet{bruderer2012}, \citet{fedele2013b} and \citet{podio2013}. %, all based on selected intermediate-$J$ CO transitions.  
% wide spec range of SPIRE; lead to motivation of study
The broad instantaneous bandwidth of the {\it Herschel} SPIRE spectrometer, covering ten consecutive CO lines of \Jup=4--13, allows the first comprehensive study of the mid-$J$ CO rotational ladder, providing important information about the gas excitation conditions between radii of typically several tens au and several hundreds au.

\begin{table*}
\caption{
Key parameters of the sources. The Herbig Ae/Be discs with and without flaring, respectively, are listed first, followed by the T~Tauri discs. Within each group, the ordering is by spectral type from early to late. 
 }
\label{t:sourceparam}
\begin{tabular}{l l @{\ } l l r l r l l }
\hline
Object name (alternative) & Spectral	& [ref.]	& Cat.$^{(a)}$	& Distance & [ref.] 	& \multicolumn{2}{l}{Disc diameter [ref.]$^{(b)}$}	& Flaring$^{(c)}$     \\
			& 	\ \ type		& 		& 			& (pc)	 &		& (au)	     & 	& (y/n)	\\
\hline
%Find group classification I (flaring) or II (non-flaring) for all Herbig targets in meeus2001, acke2010, maaskant2013
% use listings of references compiled in "Tables 1" of Meeus+ (2012), Meeus+ (2013), Sturm+ (2013)
HD~100546	& B9~Ve & [1]		& HAeBe	& 103$\pm$6 & [1]	& 700 & submm [2] & y  \\
HD~179218 	& B9e & [3]  & HAeBe		&	240$^{+70}_{-40}$ & [1]	& 240$\pm$40 & model [4] & y \\
HD~97048 (CU~Cha) & B9/A0e & [1] & HAeBe	& 180$^{+30}_{-20}$ & [1] & $\sim$750 & mid-IR [5] 	& y \\  
AB~Aur (HD~31293) & A0~Ve & [1]  & HAeBe	& 139$\pm$19	& [6]	& 900 & submm [7]	& y \\
HD~36112 (MWC~758) & A5~IVe & [3] & HAeBe	& 	205 & [6]	& $\sim$330 & submm [8] 	& y \\  % disc size inferred from 1.6 arcsec diameter of submm dust continuum from SMA in Isella+ 2010b
HD~169142 (MWC~925) & A5~Ve & [3] & HAeBe & $\sim$145 & [3]	& 470$\pm$10 & mm [9]	& y \\ % hole/gap info Quanz+ 2013b
HD~100453	& A9~Ve & [3]	& HAeBe		& 122$\pm$10 & [6]	& 	$\sim$100 & optical [10]	& y \\
HD~142527	& F6 IIIe & [11]	& HAeBe	& 145$\pm$15 & [12]	& $\sim$600 & submm [13]	& y \\
HD~50138 (MWC~158)		& B6-7~III-Ve & [14] & HAeBe$^{(d)}$		& 500$\pm$150 & [6] 	& $>$30 & mid-IR [15] & n \\ 
HD~163296 (MWC~275) & A1-3~Ve & [16] & HAeBe & 119$^{+13}_{-10}$ & [6] & $\sim$1000 & submm [17]  & n \\
HD~104237 (DX~Cha) & A4~IVe & [3] & HAeBe	& 115$\pm$5 & [6]	& -	&		& n \\
HD~144432	& A9/F0~Ve & [18]	& HAeBe	& 160$^{+36}_{-25}$ & [6] & 120$\pm$40 & model [4] & n \\  % sp.type uncertain, see refs. on p9 of Sandell+ 2011, 
RNO~90		& G5	 & [19]	& T~Tauri		& 125 & [20]	& -	&		& -  \\
RY~Tau		& K1	 & [19]	& T~Tauri		& 140 & [21]	& $\sim$140 & mm [22]		& - \\ 
DR~Tau		& K5 & [23]	& T~Tauri		& 140 & [21]	& 180--380 & (sub)mm [24]	& - \\
TW~Hya		& K7 & [25]	& T~Tauri		& 51$\pm$4 & [26]	& 400 & (sub)mm [27]	& - \\
FZ~Tau		& M0 & [28]	& T~Tauri		& 140 & [21]	& 200--600 & mm [24]		& - \\
VW~Cha		& M0.5 & [29] 	& T~Tauri		& 180 & [30]	& $\la$30 & near-IR [31]		& - \\
\hline
\end{tabular} 
\begin{minipage}{\textwidth}
\textit{Notes}. 
$^a$Category based on spectral type of central star. 
$^b$Disc diameter is taken from resolved imaging at millimetre or submillimetre wavelengths, if available; otherwise from other wavelengths or inferred from model fits. A `-' symbol indicates an unknown disc diameter. 
$^c$The division of Herbig Ae/Be objects into `flaring' and `non-flaring' (flat) disc geometries is the same as in \citet[Table~1]{meeus2013}, based on the mid-infrared dust SED shape \citep{meeus2001}. A similar classification has not yet emerged for T~Tauri objects, because their dust SEDs have a less pronounced dichotomy and they are typically more embedded in surrounding material. 
$^d$Although usually classified as a pre-main-sequence Herbig star, HD~50138 could also be an evolved B[e] star \citep[e.g.,][]{borgesfernandes2012}. 
References --- 
[1] \citet*{vandenancker1998}; 
[2] \citet{pineda2014}; 
[3]~\citet{vanboekel2005a}; 
[4] \citet[disc diameter from model fit to unresolved data]{dent2005}; 
[5] \citet{lagage2006}; 
[6] revised Hipparcos parallaxes \citep{vanleeuwen2007}; 
[7] \citet{lin2006}; 
[8] \citet{isella2010b}; 
[9] \citet{raman2006}; 
[10] \citet[who infer the disc size of HD~100453 from various observations, including scattered light images from Hubble]{collins2009}; 
[11] \citep{verhoeff2011}; 
[12]~\citet{acke2004b}; 
[13] \citet{fukagawa2013}; 
[14] \citet{borgesfernandes2009}; 
[15] \citet{borgesfernandes2011}; 
[16] \citet{tilling2012}; 
[17] \citet{isella2007}; 
[18] \citet*{dunkin1997a}; 
[19] \citet{chen1995}; 
[20] \citet*{pontoppidan2011}; 
[21] Associated with the Taurus star-forming region, typically assumed to be at a distance of 140~pc \citep{torres2009}; 
[22] \citet*{isella2010a}; 
[23] \citet{mora2001}; 
[24] \citet{ricci2010a}; 
[25] \citet{web1999}; 
[26] \citet{mamajek2005}; 
[27] \citet{qi2004}; % Rout=196AU gives good match of their submm SMA CO gas maps
[28] \citet{rebull2010}; 
[29] \citet{guenther2007}; 
[30] \citet{pontoppidan2010a}; 
[31] \citet[who infer a limit to VW Cha's disc size from near-infrared sub-arcsecond imaging]{brandeker2001}. 
\end{minipage}
\end{table*}

This paper is organized as follows. 
Section~\ref{sec:spireobs} describes the SPIRE FTS observations and the data processing. Section~\ref{sec:lineresults} presents all spectral line detections and upper limits. The detected CO rotational ladders are analysed and compared with existing, detailed physical-chemical protoplanetary disc models in Section~\ref{sec:analysis}. Section~\ref{sec:discussion} describes tentative trends in the dataset and discusses detections and upper limits of species other than CO. Finally, the most important conclusions are summarized in Section~\ref{sec:conclusions}.

% ------ obs -------
\section{{\it Herschel} SPIRE spectroscopic observations and data processing}
\label{sec:spireobs}

A sample of 18 protoplanetary discs was observed with the SPIRE spectrometer \citep{griffin2010} on-board the {\it Herschel} Space Observatory \citep{pilbratt2010} in the context of two Guaranteed Time (GT) projects and one Open Time project. Table~\ref{t:spireftsobs} outlines the observational details: twelve targets under {\tt GT1\_golofs01\_4},
%\footnote{There is a thirteenth object observed under program {\tt GT1\_golofs01\_4}: T Tau. We leave this target out of consideration here, because the continuum and line emission from this source is dominated by a protostellar envelope rather than the circumstellar disc which is the topic of this study.}
two in {\tt GT2\_jbouwman\_3} and four in {\tt OT1\_kponto01\_1}. The execution time (and on-source integration time) for the targets studied in this paper from each of the three respective programmes was 12.9 (12.0) h, 1.6 (1.5) h and 9.9 (9.5) h. 

% spectral range, point-like targets
The observations span the spectral range between 450~GHz and 1540~GHz (666--195~\micron), simultaneously observed in the two bands of the SPIRE Fourier Transform Spectrometer (FTS). The spectrometer long wavelength (SLW) and spectrometer short wavelength (SSW) bands span 450--985~GHz and 960--1540~GHz, respectively. The observed discs, with angular sizes of at most a few arcseconds, are point-like relative to the angular resolution of the {\it Herschel} SPIRE FTS (beam FWHM of 43--17\arcsec). The objects were therefore observed in sparse spatial sampling mode (see SPIRE Observers' Manual \citeyear{spireobsmanual2011}), with the central detectors of the two arrays pointed at the source and the off-centre detectors providing a measurement of diffuse, extended background emission. 
% calibration accuracy
The absolute calibration uncertainty of SPIRE FTS for point-like sources observed in sparse spatial sampling mode is 6\% \citep{swinyard2014}. 
% HR mode:
To identify any line emission present in the observed frequency range, the high spectral resolution mode was chosen \citep[see][]{spireobsmanual2011}. Lines originating from cold gas in individual circumstellar discs will be spectrally unresolved, given the instrumental resolution of 1.2~GHz ($\nu/\Delta\nu\approx400$--$1300$).

\begin{table*}
\caption{SPIRE FTS observational details. 
The 12 targets in the first segment of the table were observed as part of {\it Herschel} programme {\tt GT1\_golofs01\_4}, HD 179218 and HD 50138 were observed under programme {\tt GT2\_jbouwman\_3}, and the 4 targets in the last segment under programme {\tt OT1\_kponto01\_1}. Observation coordinates, listed in column (2) and (3), are uncertain by typically $\sim$2\arcsec\ \citep{sanchez-portal2014}. {\it Herschel} observation identifiers (obs.~id.) and observation dates are listed in columns (4) and (5). Column (6) lists on-source exposure time: 133.25~s for each scan pair, i.e., one forward and one reverse scan. Column (7) denotes whether or not any background emission was subtracted from the spectrum. Column (8) indicates whether or not a non-standard pipeline was applied to correct effects of low cryocooler evaporator (CEV) temperature. 
}
\label{t:spireftsobs}
\begin{tabular}{l r r c c c c c}
\hline
\mcol{(1)} & \mcol{(2)} & \mcol{(3)} & \mcol{(4)} & \mcol{(5)} & \mcol{(6)} & \mcol{(7)} & \mcol{(8)} \\
Object name	& \multicolumn{2}{c}{Sky coordinates}& {\it Herschel} & Observation date	& $t_\mathrm{exp}$ & background & low CEV temper- \\
	& Right Ascension 		& Declination &  Obs.~id. & & & subtracted & ature corrected \\ 
 	& (J2000) 			& (J2000)		& 		& (yyyy-mm-dd)	& (s)	& (y/n) & (y/n) \\
\hline
% exposure time on-source = 133.25 sec times numreps (1 rep = 1FWD+1REV)
% most targets 24 reps, RY Tau 34 reps; TW Hya 50 reps; 
%     HD179218 and HD50138 (both from Bouwman) 20 reps each
%     VW Cha and FZ Tau 75 reps each; RNO90 50 reps; DR Tau 58 reps
HD 100546	& 11\h33\m25\fs44	& $-$70\degr11\arcmin41\farcs2	& 1342202273 & 2010-08-08 & 3198 & y & n \\
TW Hya		& 11\h01\m51\fs91	& $-$34\degr42\arcmin17\farcs0	& 1342210862 & 2010-12-07 & 6663 & y & n \\  % 50 reps
HD 142527	& 15\h56\m41\fs89	& $-$42\degr19\arcmin23\farcs3	& 1342214821 & 2011-02-26 & 3198 & y & n\\
HD 144432	& 16\h06\m57\fs96	& $-$27\degr43\arcmin09\farcs8	& 1342214830 & 2011-02-26 & 3198 & y & n\\
RY Tau		& 4\h21\m57\fs41 	& $+$28\degr26\arcmin35\farcs6	& 1342214857 & 2011-02-28 & 4531 & y & n\\
HD 104237	& 12\h00\m05\fs08	& $-$78\degr11\arcmin34\farcs6	& 1342216876	& 2011-03-27 & 3198 & y & y \\  % 34 reps
HD 97048	$^\mathrm{a}$	& 11\h08\m03\fs32	& $-$77\degr39\arcmin17\farcs5	& 1342216877	& 2011-03-27 & 3198 & y & y \\
HD 36112		& 5\h30\m27\fs53	& $+$25\degr19\arcmin57\farcs1	& 1342216886 & 2011-03-27 & 3198 & y & n \\
AB Aur		& 4\h55\m45\fs84	& $+$30\degr33\arcmin04\farcs3 	& 1342216887	& 2011-03-27 & 3198 & y & n\\
HD 169142	& 18\h24\m29\fs78	& $-$29\degr46\arcmin49\farcs4	& 1342216904	& 2011-03-28 & 3198 & n & n\\
HD 163296	& 17\h56\m21\fs29	& $-$21\degr57\arcmin21\farcs9	& 1342216906	& 2011-03-28 & 3198 & y & n\\
HD 100453	& 11\h33\m05\fs58	& $-$54\degr19\arcmin28\farcs5 	& 1342224748	& 2011-07-25 & 3198 & n & y \\
\hline
HD 179218	& 19\h11\m11\fs25	& $+$15\degr47\arcmin15\farcs6	& 1342243607	& 2012-04-01 & 2665 & y & n\\ % 20 reps
HD 50138		& 6\h51\m33\fs40	& $-$6\degr57\arcmin59\farcs4	& 1342245118	& 2012-04-27 & 2665 & y & n\\ % 20 reps 
\hline
VW Cha		& 11\h08\m01\fs81	& $-$77\degr42\arcmin28\farcs7	& 1342224751 & 2011-07-25 & 9994 & y & y \\ % 75 repetitions
RNO 90		& 16\h34\m09\fs17	& $-$15\degr48\arcmin16\farcs8	& 1342228704 & 2011-09-17 & 6663 & y & y \\ % 50 repetitions
DR Tau		& 4\h47\m06\fs22	& $+$16\degr58\arcmin42\farcs9	& 1342228735 & 2011-09-17 & 7729 & y & n\\  % 58 repetitions
FZ Tau		& 4\h32\m31\fs76	& $+$24\degr20\arcmin03\farcs0	& 1342249471 & 2012-08-13 & 9994 & y & n\\  % 75 repetitions
\hline
\end{tabular} 
\begin{minipage}{\textwidth}
\textit{Notes}. 
$^a$The actual pointing of the HD 97048 observation was $\sim$3\arcsec\ away from the requested sky coordinates listed in columns (2) and (3). 
\end{minipage}
\end{table*}

\begin{figure*}
    \includegraphics[width=\textwidth]{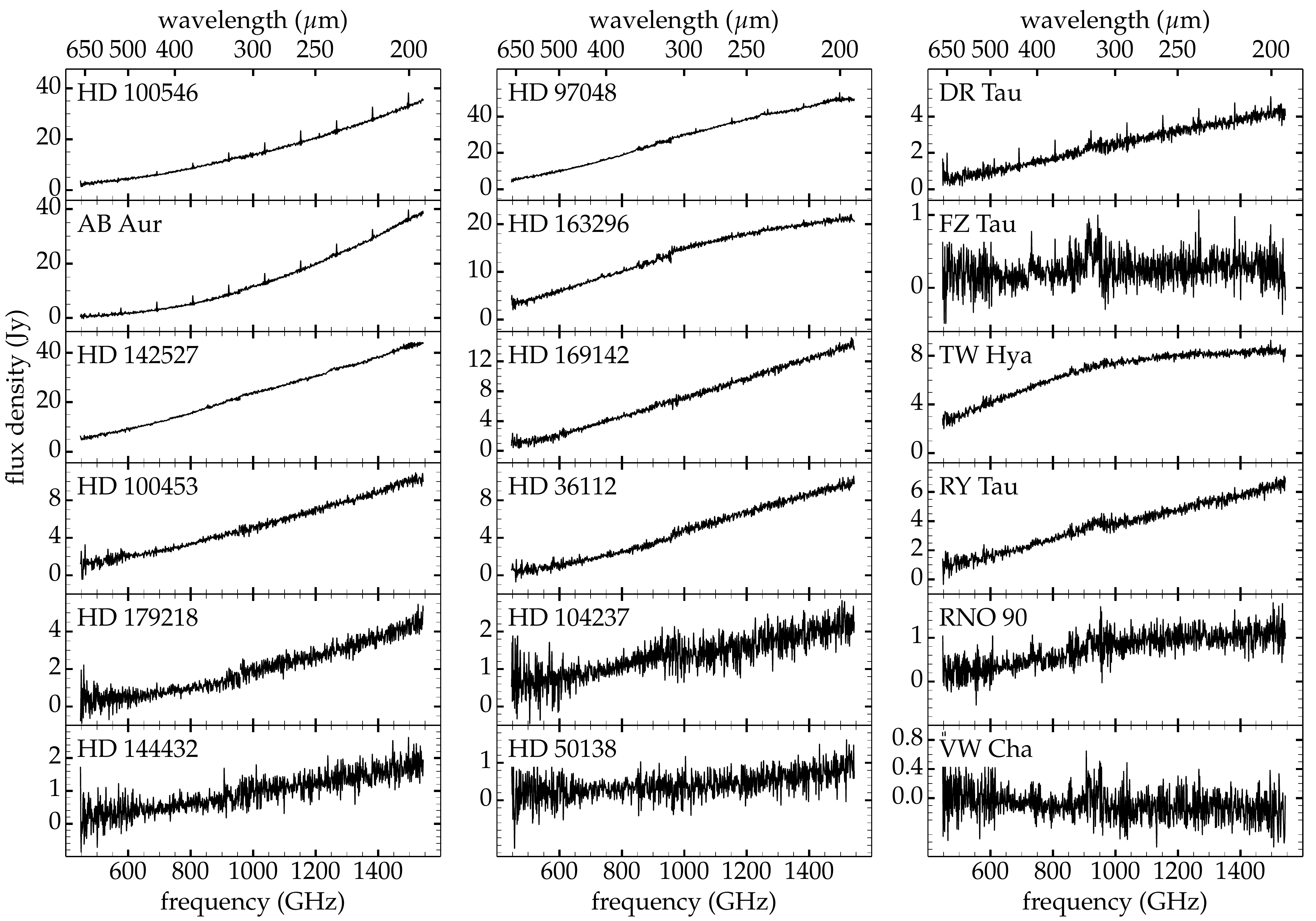}
    \caption{Background-subtracted SPIRE spectra of all sources studied in this paper. The high frequency end of SLW data ($>$960~GHz) is not shown, because they are noisier than data from the overlapping part of SSW. The line-rich sources highlighted later and in Fig.~\ref{fig:linespectra} are in the two top rows; the remaining sources are ordered by decreasing continuum intensity. T~Tauri objects are all in the rightmost column. }
    \label{fig:continuumspectra}
\end{figure*}

% promise to make highly processed products available through HSA
The processed spectra of the two GT projects have been made available in electronic form through the `user provided data products' section of the {\it Herschel} Science Archive\footnote{The user provided data products at the {\it Herschel} Science Archive are available at \url{http://www.cosmos.esa.int/web/herschel/user-provided-data-products}\,.}. The steps taken in the data processing are described in the remainder of this section. 

% data processing
The raw data were processed with the SPIRE FTS `point source' pipeline \citep{fulton2010} in version 11 of HIPE, the {\it Herschel} Interactive Processing Environment \citep{ott2010}. A non-standard version of the pipeline was used for five specific observations (see Table~\ref{t:spireftsobs}), which included an empirical correction\footnote{In future versions of HIPE, from version 13 onwards, the standard SPIRE FTS pipeline will systematically correct for low cryocooler evaporator temperature.} of the SLW continuum shape for effects of an anomalously low cryocooler evaporator temperature (Hopwood, private communication; see also \citealt{hopwood2013herschelconf}). The resulting data products contain the source spectra, observed with the central detectors of the two arrays and spectra of the surrounding background emission observed with the remaining unvignetted detectors (16 for SSW and 6 for SLW). 

% off-centre subtraction
For the majority of our source sample, a contribution from diffuse, extended background emission is evident and results in an offset in flux density between spectra from the central detectors of SLW and SSW at frequencies where the two bands overlap (960--985~GHz). If the source spectrum is contaminated by emission from spatially extended components, the considerably larger and more complex SLW beam \citep{makiwa2013} will pick up more power than the corresponding SSW detector in the same spectral range. 
To establish a representative spectrum for the background emission component, measurements from the off-centre detectors (first ring for SLW, second ring for SSW, both centred at $\sim$50\arcsec\ from the central position) are used. Data from the known outlier detectors SSWE2 and SSWF2 are always discarded, which is standard practice in the background subtraction script provided within HIPE. Additional visual inspection of data from the remaining nine SSW and six SLW detectors ensures that no additional outliers are folded into the background determination. In practice, at most one of these spectra is excluded. The `spectrometer background subtraction' script in HIPE calculates an average background spectrum, which is subsequently subtracted from the central spectrum. No significant background is detected in the observations of HD~169142 and HD~100453 and background subtraction is therefore not applied (see Table~\ref{t:spireftsobs}). For these two objects, the intensity levels measured in the frequency range where the two bands overlap already match well. For all other targets, the background subtraction strategy reduces or eliminates the intensity offset between the two bands. The resulting spectra are presented in Fig.~\ref{fig:continuumspectra}. 

% off-centre subtraction also catches diffuse line signal
The off-centre subtraction method applied here subtracts not only continuum emission originating in the diffuse background, but also spectral line signal, where present. This process ensures that contributions to the line signal from diffuse interstellar gas are subtracted before the results are interpreted in the context of circumstellar disc physics. The extended nature of specific spectral lines is discussed further in Section~\ref{sec:lineresults}.

% mis-pointing of HD97048 obs
In the particular case of the spectrum of HD~97048, the subtraction of background emission led to an increased jump in intensity between the two bands, with the SSW signal falling below that of SLW. A pointing offset of 3\farcs6 ($\pm$2\arcsec) was discovered in the observation of this target, providing a possible explanation for the jump. Therefore, after the background subtraction, the measured flux density `missed' due to the pointing offset was corrected, amounting to a frequency-dependent increase of 7--12\% across the SSW band \citep{valtchanov2014}. Since the SLW beam is considerably larger than the SSW beam, the flux loss due to a pointing error is $<$1\% across the SLW band in this case. 

%Angular extent of the discs, no SECT applied
The angular sizes of the protoplanetary disc objects studied in this paper (Table~\ref{t:sourceparam}) are generally small compared to the beam size of SPIRE FTS (17--43\arcsec). In principle, the point source pipeline calibration is only suitable for fully pointlike targets. At the upper end of the disc size estimates, the largest targets, HD~100546, HD~163296 and TW~Hya, would have an angular diameter of $\sim$8--9\arcsec. 
While a method exists to calibrate measurements of semi-extended sources with SPIRE FTS \citep{wu2013}, we choose not to scale our spectra to compensate for effects of a deviation from a fully point-like emission region, for the following reasons. Firstly, for the above cases the resulting frequency-dependent flux scale factor would be typically $<$2\% to at most 7\%. Corrections for all other targets would be even smaller. 
Secondly, caution should be exercised before applying a general `correction' based on the spatial extent of a source, since the subcomponents that give rise to emission in dust continuum and in several atomic and molecular lines may have distinctly different sizes. Such is the case for example for HD~100546, AB~Aur, HD~163296, and TW~Hya, where interferometric images show that the CO gas disc extends to radii several times larger than the dust disc (\citealt{pineda2014}; \citealt*{pietu2005}; \citealt{isella2007,degregorio-monsalvo2013,andrews2012}).

% ------- results --------
\section{Observed spectral line emission}
\label{sec:lineresults}

\begin{figure*}
    \includegraphics[width=\textwidth,angle=0]{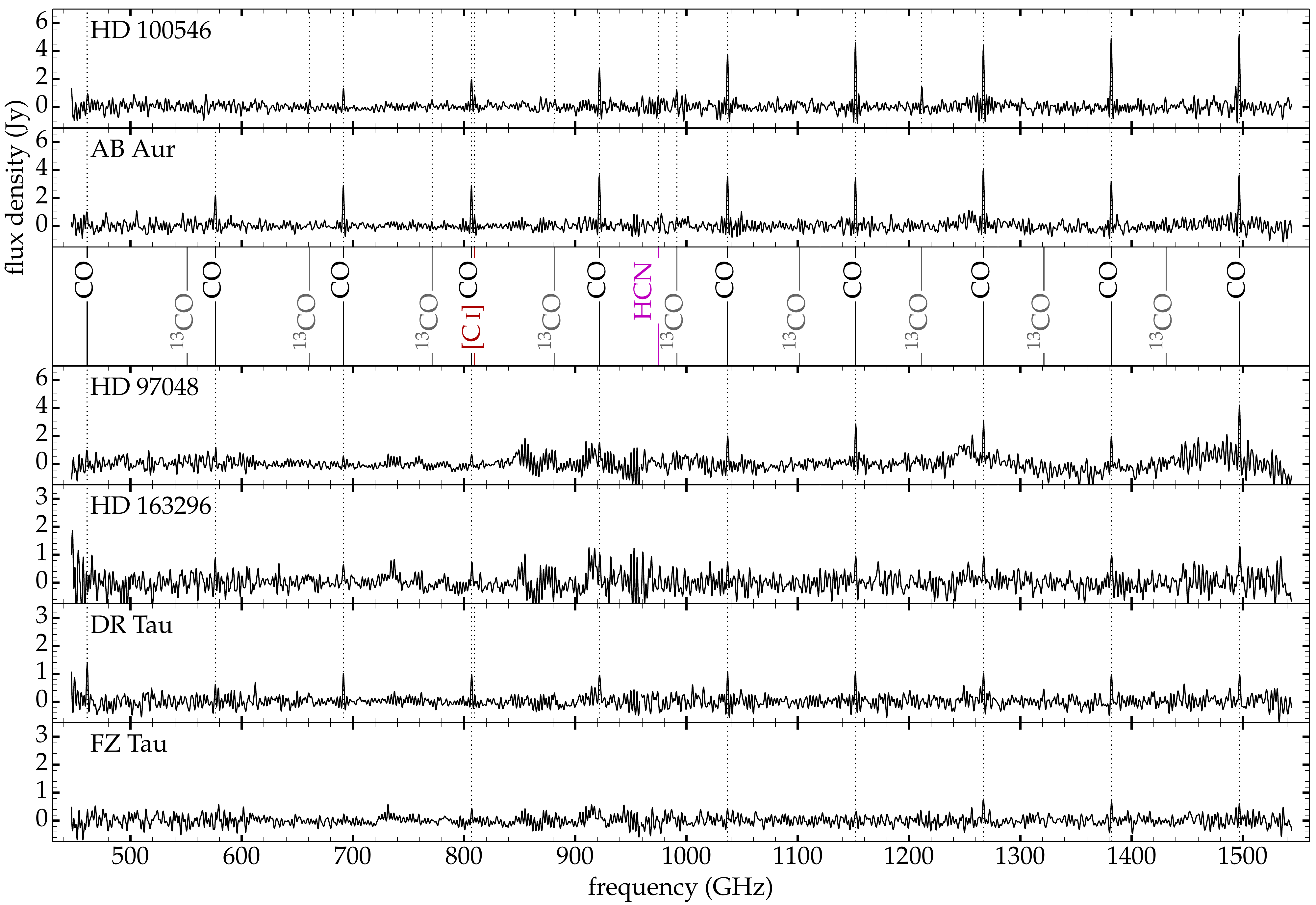}
    \caption{SPIRE spectra of sources with more than three spectral lines detected at $\geq$3$\sigma$. Spectra are shown here after subtraction of a third order polynomial fit to the continuum baseline. The high frequency end of SLW ($>$960~GHz) is not shown because it is noisier than the overlapping part of SSW. Laboratory frequencies \citep[JPL database,][]{pickett1998} for species detected in our spectra are indicated in the third panel. Dotted lines in the other panels mark those lines that are detected in each individual source. Note that each spectral line is unresolved and is dominated by the instrumental Sinc profile ({discussed in Section~\ref{sec:linedetections} and} see also Fig.~\ref{fig:bgsubzoom}), including sidelobes, mainly visible near the stronger lines, e.g., the highest frequency CO lines in the top panel. The broader spectral features visible in the HD~97048 spectrum, e.g.\ near 1480~GHz and near 1250~GHz, are not believed to be real astronomical signal. 
     }
    \label{fig:linespectra}
\end{figure*}

\subsection{Spectral line detections}
\label{sec:linedetections}

Figure~\ref{fig:continuumspectra} shows the observed dust continuum and spectral line emission for all 18 targets. A study of the multiwavelength dust spectral energy distributions of HD~100546 and HD~142527 will be presented by Bouwman et al., (in prep.) and Min et al., (in prep.). Since this study focuses on gas content of protoplanetary discs, in Fig.~\ref{fig:linespectra} we highlight specifically those six targets where more than three spectral lines are detected. 

Spectral line parameters are determined using the IDL based `FT~Fitter'\footnote{FT Fitter is developed at the University of Lethbridge and is available online at {\tt www.uleth.ca/phy/naylor/}\,.} program, which has been specifically developed to fit simultaneously both continuum and line signal from generic Fourier transform spectrometers. The continuum and lines in the spectra for each source are fit independently for the SLW and SSW bands. 
The instrumental line shape of the `unapodized' product is well represented by the canonical Sinc function. We do not apply any apodization functions, designed to mitigate the `ringing' of the Sinc lines, because they come at the expense of decreased spectral resolution \citep{naylor2007}. 

The centroid line velocities (in the respective LSR frames) derived from the fitted line centroid frequencies and the spacecraft velocity at the time of observation are always within $\sim$80~\kms\ from source velocities known from higher spectral resolution observations \citep{tang2012,dent2005,panic2008,thi2001b}. In fact, in two thirds of the cases, line velocities are consistent with known source velocities to within the associated uncertainties. Typical fitted uncertainties range from 5 to 25~\kms\ and vary inversely with the line signal-to-noise ratio. This result confirms that line centres can be determined from FTS spectra with an accuracy far exceeding the instrumental spectral line width, but that it becomes less reliable for lower signal-to-noise lines \citep{naylor2000,davis2001book}. The internal motions (turbulence, rotation) in protoplanetary discs have a magnitude of at most a few \kms\ and are therefore unresolved in our data. In this paper, we focus purely on integrated line intensities.

\begin{table*}
\caption{Integrated line intensities measured in the background-subtracted SPIRE spectra of sources with $>$3 lines detected at $\geq$3~$\sigma$. Uncertainties are listed in square brackets and represent a quadratic sum of the formal fitting error and the 1-$\sigma$ noise measured in the continuum in a 50~GHz spectral region around the line in question. For non-detections, the intensity column is marked with a `-' symbol and the uncertainty column lists the 1-$\sigma$ local continuum noise. Marginal detections ($<$3~$\sigma$) are marked by a $^*$ symbol. Rest frequencies and energy levels are taken from the JPL database \citep{pickett1998}. 
}
\label{t:spirelinefluxes}
\begin{tabular}{l r r r r@{ }l r@{ }l r@{ }l r@{ }l r@{ }l r@{ }l}
\hline
Spectral  line	& \multicolumn{2}{c}{Rest freq., wavelength} & \Eup/$k$ & \multicolumn{12}{c}{Integrated line intensity, [uncertainty]}  \\
			& 	(GHz) & (\micron) & (K) & \multicolumn{12}{c}{($10^{-18}$ \Wsqm)} \\
			& 	 	  &		&        &  \multicolumn{2}{c}{HD\,100546} & \multicolumn{2}{c}{AB\,Aur} & \multicolumn{2}{c}{HD\,97048} & \multicolumn{2}{c}{HD\,163296} & \multicolumn{2}{c}{DR\,Tau} & \multicolumn{2}{c}{FZ\,Tau}  \\
\hline
% paste output from python script below
                     CO 4--3 &     461.04 & 650.25 & 55.3 & $^*$11.3 & [5.2]$^a$  & $^*$12.9 & [5.0]$^a$  & $^*$12.9 & [5.6]  &       - & [6.9]  &    16.3 & [3.5]$^a$  &       - & [3.1]  \\
                     CO 5--4 &     576.27 & 520.23 & 83.0 &      - & [4.0]  &    24.7 & [3.8]$^a$  & $^*$11.4 & [5.3]  & $^*$10.4 & [4.0]  & $^*$7.4 & [2.5]  &       - & [2.3]  \\
                     CO 6--5 &     691.47 & 433.56 & 116.2 &    15.7 & [2.3]  &    33.0 & [2.6]$^a$  & $^*$6.3 & [3.5]  & $^*$7.4 & [2.9]  &    11.8 & [1.4]  &       - & [1.2]  \\
                     CO 7--6 &     806.65 & 371.65 & 154.9 &   23.5 & [2.7]  &    34.9 & [2.8]  & $^*$8.2 & [3.7]  & $^*$9.0 & [3.0]  &    11.6 & [1.5]  &     5.3 & [1.7]  \\
                     CO 8--7 &     921.80 & 325.23 & 199.1 &    33.1 & [3.5]  &    43.0 & [3.6]  & $^*$18.4 & [7.5]  & $^*$12.4 & [5.6]  &    11.6 & [2.3]  &       - & [2.4]  \\
                     CO 9--8 &    1036.91 & 289.12 & 248.9 &   45.3 & [4.1]  &    43.0 & [4.3]  &    23.8 & [5.4]  & $^*$9.1 & [3.8]  &    12.8 & [2.6]  & $^*$5.1 & [2.0]  \\
                    CO 10--9 &    1151.99 & 260.24 & 304.2 &   55.3 & [3.7]  &    41.1 & [4.0]  &    34.6 & [4.7]  &    11.7 & [3.4]  &    12.4 & [2.2]  & $^*$3.9 & [1.8]  \\
                   CO 11--10 &    1267.01 & 236.61 & 365.0 &   51.3 & [4.5]  &    48.6 & [4.5]  &    36.4 & [6.8]  &    11.3 & [2.9]  &    12.3 & [2.4]  &     9.1 & [2.1]  \\
                   CO 12--11 &    1382.00 & 216.93 & 431.3 &    58.3 & [3.3]  &    38.3 & [3.8]  &    27.9 & [5.9]  &    11.7 & [3.4]  &    11.6 & [2.1]  &     8.0 & [2.0]  \\
                   CO 13--12 &    1496.92 & 200.27 & 503.1 &    60.7 & [4.7]  &    43.0 & [4.7]  &    49.2 & [8.7]  &    15.2 & [3.7]  &    11.6 & [2.3]  &     7.5 & [2.3]  \\
              $^{13}$CO 6--5 &     661.07 & 453.50 & 111.1 & $^*$5.9 & [2.5]  &       - & [2.3]  &       - & [2.2]  &       - & [1.9]  &       - & [1.4]  &       - & [1.3]  \\
              $^{13}$CO 7--6 &     771.18 & 388.74 & 148.1 & $^*$5.8 & [2.4]  & $^*$3.8 & [2.5]  &       - & [2.7]  &       - & [2.3]  &       - & [1.3]  &       - & [1.1]  \\
              $^{13}$CO 8--7 &     881.27 & 340.18 & 190.4 & $^*$6.3 & [3.4]  &       - & [3.1]  &       - & [5.6]  &       - & [4.0]  &       - & [1.8]  &       - & [2.0]  \\
              $^{13}$CO 9--8 &     991.33 & 302.41 & 237.9 & $^*$13.9 & [4.8]  & $^*$7.1 & [4.0]  &       - & [4.3]  &       - & [3.5]  &       - & [2.7]  &       - & [2.4]  \\
            $^{13}$CO 11--10 &    1211.33 & 247.49 & 348.9 &    16.8 & [3.4]  &       - & [3.0]  &       - & [4.2]  &       - & [3.0]  &       - & [1.9]  &       - & [1.7]  \\
 \Ci~$^3$P$_2$--$^3$P$_1$ &     809.34 & 370.41 & 62.5 & $^*$7.6 & [2.7]  & $^*$5.6 & [2.8]  &       - & [2.3]  &       - & [1.9]  & $^*$1.4 & [1.5]  &       - & [1.2]  \\
                  HCN 11--10 &     974.49 & 307.64 & 974.5 & $^*$8.9 & [5.5]  & $^*$6.4 & [4.2]  &       - & [4.9]  &       - & [4.5]  &       - & [2.5]  &       - & [2.5]  \\
%              (unknown)649.8 &     649.81 &         - & [2]  &         - & [2]  &         - & [2]  &         - & [2]  &         4 & [1]  &         - & [1]  \\
%              (unknown)731.6 &     731.58 &         - & [2]  &         - & [2]  &         - & [2]  &         - & [2]  &         - & [1]  &         7 & [2]  \\                  
%             (unknown)1447.4 &    1447.40 &         - & [4]  &         - & [4]  &         - & [8]  &         - & [3]  &         7 & [2]  &         - & [2]  \\
\hline
\Ci~$^3$P$_1$--$^3$P$_0$ &     492.16 & 609.14 & 23.6 &       - & [4.3]  &       - & [4.2]  &       - & [4.0]  &       - & [3.7]  &       - & [2.5]  &       - & [2.4]  \\
               \CHplus\ 1--0 &     835.08 & 359.00 & 40.1 &         - & [2.1]  &         - & [2.3]  &         - & [5.3]  &         - & [3.2]  &         - & [1.4]  &         - & [1.6]  \\
  \water\ 1$_{10}$--1$_{01}$ &     556.94 & 538.29 & 61.0 &         - & [3.8]  &         - & [3.8]  &         - & [4.2]  &         - & [3.1]  &         - & [2.0]  &         - & [2.4]  \\
  \water\ 2$_{11}$--2$_{02}$ &     752.03 & 398.64 & 136.9 &        - & [2.1]  &         - & [2.0]  &         - & [3.2]  &         - & [3.0]  &         - & [1.5]  &         - & [1.5]  \\
  \water\ 2$_{02}$--1$_{11}$ &     987.93 & 303.46 & 100.8 &        - & [4.9]  &         - & [3.7]  &         - & [4.3]  &         - & [3.5]  &         - & [2.7]  &         - & [2.3]  \\
  \water\ 3$_{12}$--3$_{03}$ &    1097.36 & 273.19 & 249.4 &       - & [3.2]  &         - & [2.8]  &         - & [2.9]  &         - & [2.4]  &         - & [1.5]  &         - & [1.4]  \\
  \water\ 1$_{11}$--0$_{00}$ &    1113.34 & 269.27 & 53.4 &         - & [3.1]  &         - & [2.7]  &         - & [2.6]  &         - & [2.9]  &         - & [1.5]  &         - & [1.5]  \\
  \water\ 3$_{12}$--2$_{21}$ &    1153.13 & 259.98 & 249.4 &       - & [2.9]  &         - & [3.7]  &         - & [3.2]  &         - & [3.1]  &         - & [1.8]  &         - & [1.6]  \\
  \water\ 3$_{21}$--3$_{12}$ &    1162.91 & 257.79 & 305.2 &       - & [2.4]  &         - & [4.0]  &         - & [3.0]  &         - & [3.1]  &         - & [2.2]  &         - & [1.7]  \\
  \water\ 2$_{20}$--2$_{11}$ &    1228.79 & 243.97 & 195.9 &       - & [4.1]  &         - & [3.8]  &         - & [6.2]  &         - & [3.5]  &         - & [2.1]  &         - & [1.9]  \\
% end of script output paste area
\hline 
\end{tabular} \\
\begin{minipage}{\textwidth}
\textit{Notes.} $^a$Suspected contribution from a marginally extended envelope.
\end{minipage}
\end{table*}

\begin{figure*}
    \includegraphics[width=1.0\textwidth,angle=0]{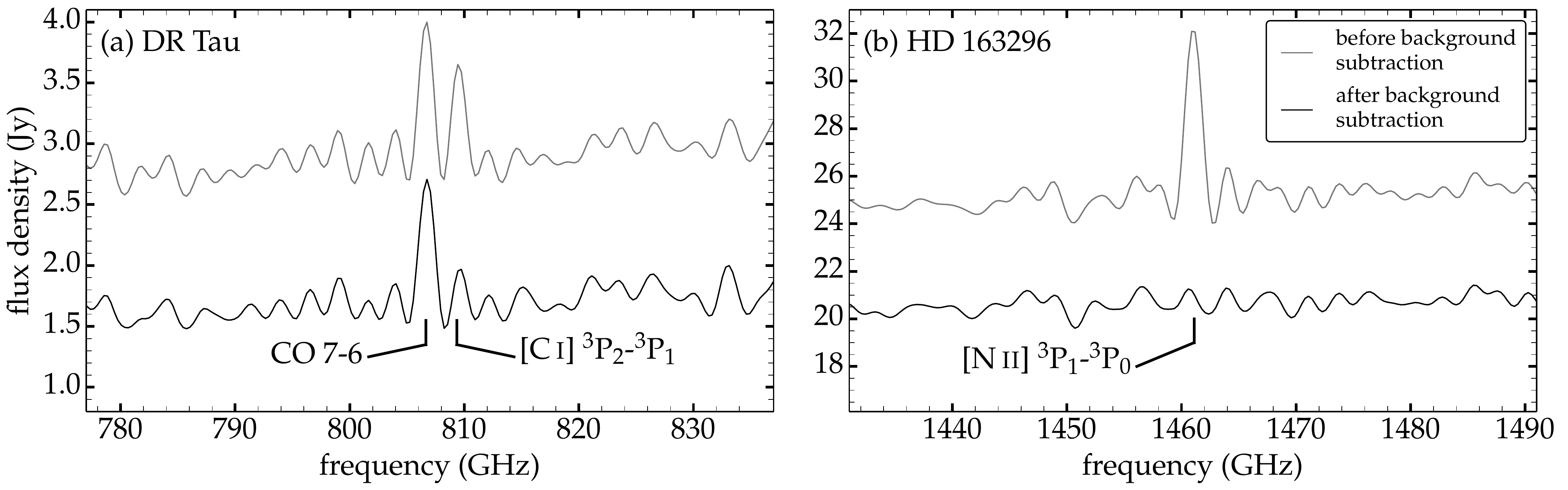}
    \caption{(a) Zoom of the 60 GHz section of the DR Tau spectrum centred around the CO 7--6 and \Ci\ $^3$P$_2$--$^3$P$_1$ lines. The line of the neutral carbon atom coincides with the first sidelobe of the neighbouring CO line, but has significant flux of its own before background subtraction (grey), while its strength is reduced considerably after background subtraction is applied (black; same spectrum as in Fig.~\ref{fig:linespectra}). (b) Zoom of the 60 GHz section of the HD~163296 spectrum centred around the \Nii\ line, again before and after background subtraction. This shows that the \Nplus\ line towards HD~163296 originates from the extended background medium.}
    \label{fig:bgsubzoom}
\end{figure*}

In Table~\ref{t:spirelinefluxes}, we present the integrated intensities of lines detected in the spectra shown in Fig.~\ref{fig:linespectra}, i.e., for sources with $>$3 lines detected at $\geq$3~$\sigma$. Integrated intensities of lines from sources with fewer highly significant detections are presented in Section~\ref{sec:upperlimits} and Table~\ref{t:upperlimits}, along with upper limits for non-detections. The large majority of line detections are identified as pure rotational lines of $^{12}$CO and \thCO. There are a handful of unidentified line detections, but all are below the 4~$\sigma$ level and therefore not listed in Tables~\ref{t:spirelinefluxes} and \ref{t:upperlimits}. 

All integrated intensities given in Table~\ref{t:spirelinefluxes} represent line signal associated with the compact disc objects, corrected for contributions from diffuse, extended background. The information from off-centre detectors in the SPIRE FTS detector arrays allows us to separate spatial components of emission (see Section~\ref{sec:spireobs}). 
For example, line emission from the fine structure transitions of the neutral carbon atom (C$^0$) at 492 GHz and 809 GHz is seen towards all objects that exhibit CO emission. Due to comparable line strengths at off-centre positions around most targets, however, these C$^0$ signatures disappear after background subtraction has been applied. Figure~\ref{fig:bgsubzoom}a shows zoomed spectra of DR~Tau before and after background subtraction. The original CO 7--6 line signal is not affected by a contribution from diffuse background and is purely attributable to compact emission. On the other hand, only a small fraction of the original flux in the neighbouring \Ci\ line originates from the compact object. This situation is seen not only for DR~Tau, but for all targets with significant line emission, where often both \Ci\ fine structure lines ($^3$P$_1$--$^3$P$_0$ and $^3$P$_2$--$^3$P$_1$) are detected, but little to nothing remains after subtracting the extended background contribution. Similarly, the strong \Nplus~$^3$P$_1$--$^3$P$_0$ fine structure line detected at 1461 GHz towards HD~163296 (Fig.~\ref{fig:bgsubzoom}b) is found to originate exclusively in surrounding background medium. The two examples shown in Fig.~\ref{fig:bgsubzoom} serve to illustrate the importance of background subtraction of both continuum and line signatures, to avoid misinterpretations of the data. 
Finally, the lowest $J$ CO lines that we detect also include significant contributions from extended emission, e.g., in AB~Aur, where, again, the off-centre detectors allow a subtraction of the diffuse background. 

% some slightly spatially extended could be missed by our off-centre detectors
While the background subtraction method applied here accounts for truly diffuse emission with angular scales of $\ga$1\arcmin, an emission region that is less extended, but still larger than pointlike, i.e., $\sim$10--30\arcsec, could remain unseen by the ring of off-centre detectors used to probe extended emission. Such angular scales cover sizes of protostellar envelopes that are known to surround some of the star-disc systems in our sample. We suspect contributions from an extended envelope in at least the following cases: the few lowest energy SPIRE CO lines in AB~Aur (cf.~spectrally resolved components in CO 2--1 in \citealt{tang2012}), CO~4--3 in DR~Tau (cf.~CO 3--2 in \citealt{thi2001b}) and possibly HD~100546 (cf.~non-detection of CO~5--4 in Table~\ref{t:spirelinefluxes}).  

% == upper limits ==
\subsection{Upper limits and faint detections}
\label{sec:upperlimits}

\begin{table*}
\caption{
Integrated intensities of (weak) lines in the 12 targets not listed in Table~\ref{t:spirelinefluxes}. Uncertainties, in square brackets, are the same type of 1-$\sigma$ errors as in Table~\ref{t:spirelinefluxes}. Most entries in this table are non-detections, marked by a `-' symbol; in these cases the uncertainty represents the 1-$\sigma$ upper limit. The top section contains lines that are detected in at least one source (see Table~\ref{t:spirelinefluxes}); the bottom section contains limits for lines that are not detected in any sources. Rest wavelengths and upper level energies for each transition are given in Table~\ref{t:spirelinefluxes}. 
}
\label{t:upperlimits}
\rotatebox{90}{ % to rotate entire table
\begin{tabular}{l @{} r @{\ \ } r@{ }l @{\ \ } r@{ }l @{\ \ } r@{ }l r@{ }l @{\ \ } r@{ }l r@{ }l @{\ \ } r@{ }l @{\ \ } r@{ }l @{\ \ } r@{ }l r@{ }l r@{ }l r@{ }l}
\hline
Spectral 		& Rest freq.	& \multicolumn{24}{c}{Integrated line intensity, [uncertainty]}  \\
line			& 	(GHz)		& \multicolumn{24}{c}{($10^{-18}$ \Wsqm)} \\
 & & \multicolumn{2}{c}{\rotatebox{-90}{HD\,142527}} & \multicolumn{2}{c}{\rotatebox{-90}{HD\,144432}} & \multicolumn{2}{c}{\rotatebox{-90}{HD\,104237}} & \multicolumn{2}{c}{\rotatebox{-90}{HD\,36112}} & \multicolumn{2}{c}{\rotatebox{-90}{HD\,169142}} & \multicolumn{2}{c}{\rotatebox{-90}{HD\,100453}} & \multicolumn{2}{c}{\rotatebox{-90}{TW Hya}} & \multicolumn{2}{c}{\rotatebox{-90}{RY Tau}} & \multicolumn{2}{c}{\rotatebox{-90}{HD\,179218}} & \multicolumn{2}{c}{\rotatebox{-90}{HD\,50138}} & \multicolumn{2}{c}{\rotatebox{-90}{VW Cha}} & \multicolumn{2}{c}{\rotatebox{-90}{RNO 90}} \\
\hline
% paste output from python script below
                     CO 4--3 &     461.04 &       - & [5.0]  &       - & [5.3]  &       - & [6.1]  &       - & [4.5]  &       - & [5.4]  &       - & [6.5]  &       - & [4.4]  &       - & [4.7]  &       - & [7.5]  &       - & [6.2]  &       - & [3.5]  &       - & [3.0]  \\
                     CO 5--4 &     576.27 & - & [4.3]  &       - & [3.4]  & $^*$10.1 & [4.4]  &       - & [3.7]  &       - & [3.3]  &       - & [3.8]  &       - & [2.5]  &       - & [2.8]  &       - & [3.4]  &       - & [3.4]  &       - & [2.4]  &       - & [2.9]  \\
                     CO 6--5 &     691.47 &       - & [2.2]  &       - & [1.6]  &       - & [1.7]  &       - & [1.9]  & $^*$4.1 & [2.3]  &       - & [2.1]  &       - & [1.3]  &       - & [1.5]  & $^*$6.8 & [2.6]  &       - & [1.8]  &       - & [0.9]  &       - & [1.4]  \\
                     CO 7--6 &     806.65 & $^*$6.0 & [2.8]  &       - & [1.7]  & $^*$5.2 & [2.2]  &       - & [1.7]  & $^*$6.2 & [2.4]  &       - & [1.3]  &       - & [1.0]  &       - & [1.3]  &       - & [1.7]  &       - & [1.9]  &       - & [1.1]  &       - & [1.5]  \\
                     CO 8--7 &     921.80 &       - & [3.7]  &       - & [3.0]  &       - & [2.9]  &       - & [2.1]  &       - & [2.7]  &       - & [1.9]  &       - & [2.4]  &       - & [2.6]  &       - & [2.9]  &       - & [2.7]  &       - & [2.4]  &       - & [2.3]  \\
                     CO 9--8 &    1036.91 &       - & [3.4]  &       - & [2.7]  &       - & [3.0]  &       - & [3.2]  &       - & [3.0]  &       - & [3.1]  &       - & [1.8]  &       - & [2.4]  &       - & [2.9]  &       - & [2.9]  &       - & [2.2]  &       - & [2.2]  \\
                    CO 10--9 &    1151.99 &       - & [4.1]  &       - & [2.3]  &       - & [3.4]  &       - & [2.5]  &     9.5 & [2.6]  &       - & [2.3]  &       - & [1.9]  &       - & [2.2]  &       - & [2.7]  &       - & [1.9]  &       - & [2.0]  &       - & [2.2]  \\
                   CO 11--10 &    1267.01 &       - & [4.2]  &       - & [3.0]  & $^*$9.0 & [3.4]  &       - & [2.7]  &       - & [3.1]  &       - & [2.1]  &       - & [1.9]  &       - & [2.1]  &       - & [2.6]  &       - & [2.6]  &       - & [2.0]  &       - & [2.0]  \\
                   CO 12--11 &    1382.00 &       - & [3.9]  &       - & [2.4]  & $^*$7.5 & [3.2]  &       - & [2.4]  &       - & [2.6]  &       - & [2.4]  &       - & [1.6]  &     7.5 & [2.4]  &       - & [3.2]  &       - & [2.8]  &       - & [1.7]  &       - & [1.8]  \\
                   CO 13--12 &    1496.92 &       - & [7.4]  &    10.4 & [3.2]  &       - & [3.2]  &       - & [3.1]  & $^*$9.8 & [3.6]  &       - & [3.4]  &    10.7 & [2.5]  &       - & [2.8]  &       - & [3.2]  &       - & [3.5]  &       - & [2.3]  &       - & [2.8]  \\
\Ci~$^3$P$_2$--$^3$P$_1$ &     809.34 & $^*$4.2 & [2.8]  &       - & [1.7]  &       - & [1.5]  &       - & [1.7]  &       - & [1.7]  &       - & [1.3]  &       - & [1.0]  &       - & [1.3]  &       - & [1.7]  &       - & [1.9]  &       - & [1.1]  &       - & [1.5]  \\
                   \CeiO~9--8 &     987.56 &       - & [3.5]  &       - & [3.5]  &       - & [3.9]  &       - & [3.5]  &       - & [3.7]  &       - & [3.0]  &       - & [2.3]  & $^*$9.4 & [3.4]$^a$  &       - & [3.5]  &       - & [3.7]  &       - & [1.7]  &       - & [2.8]  \\
                  HCN 11--10 &     974.49 &       - & [3.1]  &       - & [3.6]  &       - & [4.3]  &       - & [3.6]  & $^*$10.3 & [4.0]  &       - & [3.6]  &       - & [2.6]  &       - & [3.7]  &       - & [3.5]  &       - & [3.8]  &       - & [1.7]  &       - & [2.9]  \\
\hline
  \Ci~$^3$P$_1$--$^3$P$_0$ &     492.16 &       - & [4.6]  &       - & [4.1]  &       - & [4.2]  &       - & [4.1]  &       - & [4.0]  &       - & [4.0]  &       - & [2.7]  &       - & [3.7]  &       - & [5.2]  &       - & [4.2]  &       - & [2.4]  &       - & [2.7]  \\
               \CHplus\ 1--0 &     835.08 &       - & [3.2]  &       - & [1.9]  &       - & [1.6]  &       - & [2.0]  &       - & [1.9]  &       - & [1.6]  &       - & [1.6]  &       - & [2.1]  &       - & [2.1]  &       - & [2.5]  &       - & [1.2]  &       - & [2.0]  \\
  \water\ 1$_{10}$--1$_{01}$ &     556.94 &       - & [3.2]  &       - & [3.1]  &       - & [4.5]  &       - & [3.3]  &       - & [3.2]  &       - & [5.0]  &       - & [2.3]  &       - & [3.0]  &       - & [4.6]  &       - & [3.0]  &       - & [2.2]  &       - & [2.5]  \\
  \water\ 2$_{11}$--2$_{02}$ &     752.03 &       - & [2.4]  &       - & [2.1]  &       - & [1.3]  &       - & [1.9]  &       - & [1.6]  &       - & [1.7]  &       - & [1.5]  &       - & [1.6]  &       - & [1.9]  &       - & [1.5]  &       - & [1.6]  &       - & [2.3]  \\
  \water\ 2$_{02}$--1$_{11}$ &     987.93 &       - & [3.5]  &       - & [3.5]  &       - & [3.9]  &       - & [3.5]  &       - & [3.7]  &       - & [3.0]  &       - & [2.3]  &       - & [3.2]$^a$  &       - & [3.5]  &       - & [3.7]  &       - & [1.7]  &       - & [2.8]  \\
  \water\ 3$_{12}$--3$_{03}$ &    1097.36 &       - & [2.8]  &       - & [2.1]  &       - & [2.9]  &       - & [3.1]  &       - & [2.8]  &       - & [2.1]  &       - & [1.7]  &       - & [2.3]  &       - & [3.3]  &       - & [2.4]  &       - & [1.5]  &       - & [2.0]  \\
  \water\ 1$_{11}$--0$_{00}$ &    1113.34 &       - & [3.2]  &       - & [2.3]  &       - & [3.3]  &       - & [3.0]  &       - & [2.7]  &       - & [2.4]  &       - & [1.7]  &       - & [2.6]  &       - & [3.4]  &       - & [2.4]  &       - & [1.8]  &       - & [2.0]  \\
  \water\ 3$_{12}$--2$_{21}$ &    1153.13 &       - & [4.3]  &       - & [2.4]  &       - & [3.2]  &       - & [2.4]  &       - & [2.2]  &       - & [2.3]  &       - & [1.9]  &       - & [1.9]  &       - & [2.8]  &       - & [1.9]  &       - & [2.1]  &       - & [2.2]  \\
  \water\ 3$_{21}$--3$_{12}$ &    1162.91 &       - & [4.6]  &       - & [2.7]  &       - & [3.6]  &       - & [2.5]  &       - & [2.3]  &       - & [2.1]  &       - & [1.7]  &       - & [1.6]  &       - & [2.9]  &       - & [2.7]  &       - & [1.8]  &       - & [2.1]  \\
  \water\ 2$_{20}$--2$_{11}$ &    1228.79 &       - & [4.6]  &       - & [2.7]  &       - & [2.9]  &       - & [2.8]  &       - & [2.9]  &       - & [2.8]  &       - & [1.9]  &       - & [2.2]  &       - & [2.6]  &       - & [2.5]  &       - & [1.8]  &       - & [2.2]  \\
% end of script pasting area
\hline
\end{tabular} 
} % end of rotatebox
\begin{minipage}{\textwidth}
\textit{Notes}. 
$^a$Likely a line blend. The intensity listed for \CeiO~9--8 in RY~Tau is unlikely to arise purely due to that line transition, given that none of the other \CeiO\ lines are detected. Instead, it may be largely due to the \water\ line at 987.9 GHz. 
\end{minipage}
\end{table*}

Table~\ref{t:upperlimits} lists line detections and upper limits in the 12 sources with up to 3 $\geq$3~$\sigma$ line detections, beyond the 6 line-rich sources already presented in Section~\ref{sec:linedetections} (see Fig.~\ref{fig:linespectra} and Table~\ref{t:spirelinefluxes}). Upper limits are calculated from local continuum rms values (in Jy) measured in 50~GHz bins, multiplied by the instrumental line width of 1.18~GHz. 
%In addition to upper limits for lines that \emph{are} detected in the sources in Table~\ref{t:spirelinefluxes}, 
Even though they are not detected in any of the spectra presented in this paper, we include limits for the \CHplus~$J$=1--0 transition and eight \water\ lines with \Eup/$k$~$<$~400~K in the bottom sections of Tables~\ref{t:spirelinefluxes} and \ref{t:upperlimits}. These important tracers are expected to provide constraints for astrochemical models (see also Section~\ref{sec:discussCHplus} and \ref{sec:discusswater}). Any spectral lines that are not listed can be assumed to have intensity limits comparable to those of nearby lines in the same spectrum. For lines at frequencies where the two bands overlap (960--985 GHz), the limit obtained from the SSW band (higher frequencies) will provide the most stringent constraint. 

As expected, noise levels in line-free continuum are generally lowest for the longest observations, i.e., those of FZ~Tau and VW~Cha (2.8~h), followed by DR~Tau (2.1~h). This increased sensitivity translates into lower 1-$\sigma$ upper limits on non-detected lines in these spectra (Tables~\ref{t:spirelinefluxes} and \ref{t:upperlimits}), compared to most other targets that were observed for $<$2~h and typically 0.9~h each (see Table~\ref{t:spireftsobs}).

% == comparison with heterodyne measurement ==
\subsection{Assessment of calibration accuracy}
\label{sec:spire-heterodyne}

For HD~100546, several spectral lines detected in our {\it Herschel} SPIRE observation (Table~\ref{t:spirelinefluxes}) have also been observed with other instruments, namely CO 6--5, CO 7--6 and \mbox{\Ci\ $^3$P$_2$--$^3$P$_1$} with CHAMP$^+$ on APEX \citep{panic2010} and CO 10--9 with HIFI on {\it Herschel} \citep{fedele2013b}. 

% CO 6--5 from APEX is consistent with SPIRE
The integrated line intensity from the APEX measurement of \mbox{CO~6--5} at 691.5 GHz, \intintens\ = 17.7$\pm$0.9 \Kkms, corresponds to (14.3$\pm$0.7)$\times$$10^{-18}$ \Wsqm, using the conversion in Appendix~\ref{sec:litdata}. This value is consistent with the SPIRE measurement of (15.7$\pm$2.3)$\times$$10^{-18}$ \Wsqm\ to within errors, even without taking into account the absolute calibration uncertainties of both instruments. 

% CO 7-6 and C 809 GHz from APEX inconsistent with SPIRE
In contrast to CO 6--5, the CO 7--6 and \mbox{\Ci~$^3$P$_2$--$^3$P$_1$} lines at 806.7 GHz and 809.3 GHz, respectively, appear significantly brighter in the SPIRE measurement than in the APEX data. The CO 7--6 line intensity from SPIRE is (23.5$\pm$2.7)$\times$$10^{-18}$~\Wsqm, roughly two times larger than the value of (11.0$\pm$1.6)$\times$$10^{-18}$~\Wsqm\ found by \citet{panic2010} with APEX. The same authors derive an upper limit of \pow{0.85}{18}~\Wsqm\ for the \Ci $^3$P$_2$--$^3$P$_1$ line, for which we find an integrated line intensity of (7.6$\pm$2.7)$\times$$10^{-18}$ \Wsqm\ from SPIRE. 
Even when the 30\% absolute calibration uncertainty for CHAMP$^+$ at these frequencies \citep{panic2010} are incorporated, their values remain inconsistent with the SPIRE measurements. The latter should be seen as more reliable, given the higher signal-to-noise and smaller calibration errors. The background subtraction for SPIRE is based on a full ring of detectors, instead of a single off-position, as used for APEX, and the result is robust against filtering certain detectors. In addition, the SPIRE CO 7--6 line flux fits in the natural trend set by the nearest $J$ CO rotational lines (see Table~\ref{t:spirelinefluxes}), which would be broken by the much lower APEX value. This internal consistency also imparts confidence in the SPIRE line detection of the nearby neutral carbon line. Possible implications of the revised neutral carbon emission line strength are discussed below in Section~\ref{sec:carbonHD100546}. 

% CO 10-9 HIFI vs. SPIRE, ~ok
The {\it Herschel} HIFI line strength of CO 10--9 reported by \citet{fedele2013b} is 3.0$\pm$0.09 \Kkms\ (\intintens), corresponding to (45$\pm$2)$\times$$10^{-18}$ \Wsqm. Compared to the (55$\pm$4)$\times$$10^{-18}$ \Wsqm\ measured by SPIRE, the two independent measurements are consistent when the calibration uncertainties of SPIRE (6\%, \citealt{swinyard2014}) and HIFI band 5a \citep[$\sim$14\%,][]{roelfsema2012} are taken into account.

% -----------------------
\section{Analysis}
\label{sec:analysis}

\subsection{Temperature components in rotation diagrams}
\label{sec:rotatdiag}

\begin{figure}
    \includegraphics[width=0.50\textwidth]{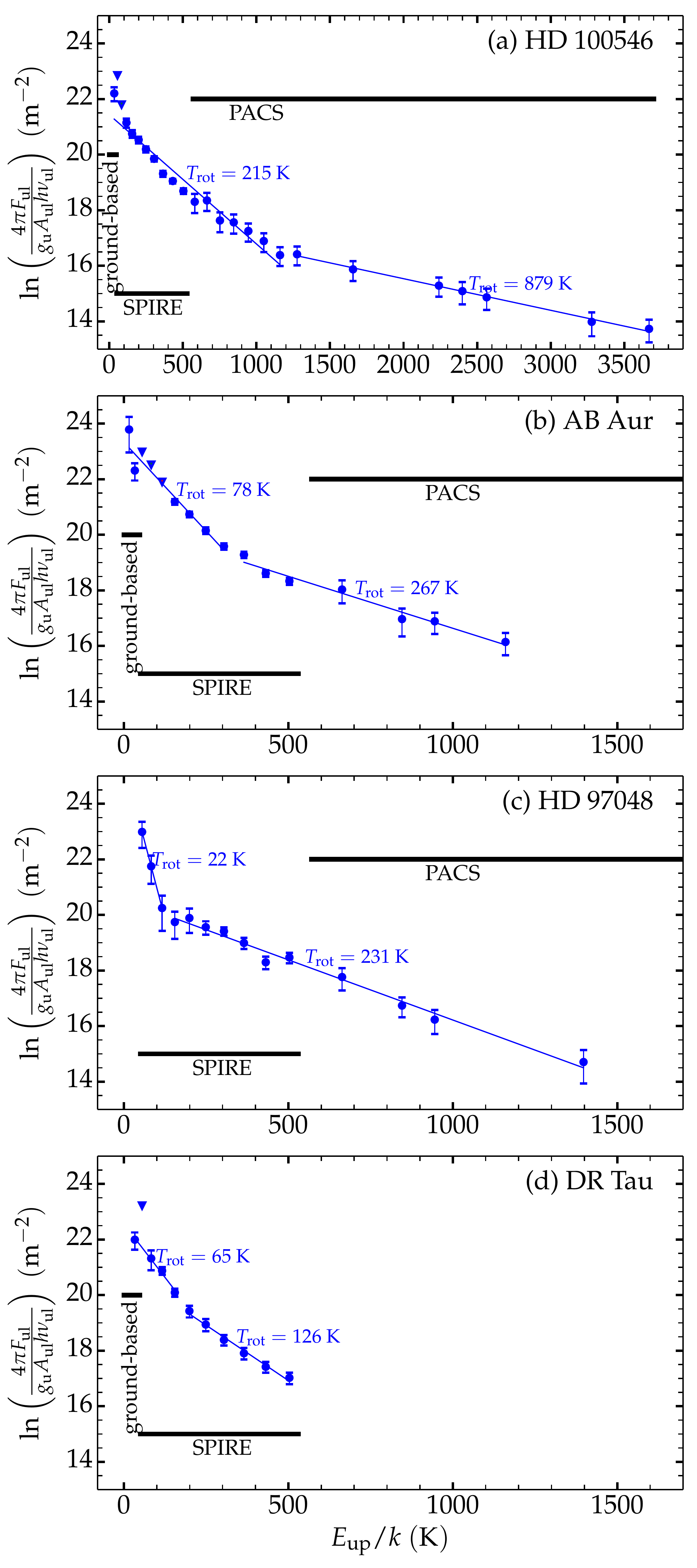} 
    \caption{Rotation diagrams for the targets with at least nine CO detections in the SPIRE and PACS bands combined. Note the different horizontal scale limits for HD~100546. Vertical error bars include contributions from measurement uncertainties (cf.~Table~\ref{t:spirelinefluxes}) and absolute calibration uncertainties. Upper limits are plotted as filled triangles, i.e., 3~$\sigma$ limits for CO 5--4 for HD~100546 and the contaminated lines for HD~100546, AB~Aur and DR~Tau (see Section~\ref{sec:linedetections} and Table~\ref{t:spirelinefluxes}). }
    \label{fig:rotatdiag}
\end{figure}

A commonly used technique to analyse a wide range of rotational transition lines from a single species is a rotation diagram \citep{goldsmith1999}. In principle, this method allows a derivation of the number of emitting molecules and of the gas temperature, but it relies on the assumptions that (i) local thermodynamic equilibrium (LTE) applies, (ii) the emitting areas are equal (or much smaller than the beam) for all observed transitions, (iii) all probed gas can be characterized by a single temperature, and (iv) the emission lines are optically thin. For pure rotational line transitions of CO in protoplanetary discs, typically only the first of these conditions holds. Nevertheless, the rotation diagram provides a useful tool to gauge variations between different disc sources. 

The vertical and horizontal axes $y$ (a function of the integrated line intensity $F_\mathrm{ul}$ for each transition from upper level `u' to lower level `l'), and $x$ (upper level energy of that transition) can be related through: 
\begin{equation}
\underbrace{ \ln \left( \frac{ 4\pi F_\mathrm{ul} }{g_\mathrm{u} A_\mathrm{ul} h \nu_\mathrm{ul}} \right)}_{\textstyle y} = \underbrace{ \vphantom{ \left( \frac{F}{g} \right) } -\frac{1}{T_\mathrm{rot}}}_{\textstyle a} \underbrace{ \vphantom{ \left( \frac{F}{g} \right) } \frac{E_\mathrm{up}}{k}}_{\textstyle x} +  \underbrace{ \ln \left( \frac{f \, \mathcal{N}}{Q \, d^2} \right) }_{\textstyle c} , 
\label{eq:rotdiag}
\end{equation}
where $h$ and $k$ are the Planck and Boltzmann constants, respectively, $g_\mathrm{u}$ is the statistical weight of state `u', $A_\mathrm{ul}$ is the Einstein coefficient for spontaneous emission, $\nu_\mathrm{ul}$ is the frequency related to the energy difference between states `u' and `l', $Q$ is the partition function, and $d$ is the distance to the source. If all the above assumptions hold, a determination of $a$ will yield the rotational temperature \Trot. Likewise, $c$ would be a measure for the total number of emitting molecules $\mathcal{N}$, but only if the beam filling factor $f$ in Eq.~(\ref{eq:rotdiag}) is either assumed to be uniform or can be constrained using high angular resolution observations. Since this paper does neither, the analysis in this section is restricted to the slope, $a$, of the rotation diagrams. 

Figures~\ref{fig:rotatdiag}a,b,c show rotation diagrams of three Herbig Ae/Be sources with at least nine CO line detections in {\it Herschel} SPIRE (this work) and {\it Herschel} PACS \citep{meeus2013} combined, augmented by ground-based datapoints in the (sub)millimetre gathered from the literature (see Appendix~\ref{sec:litdata}). Note that some of the CO rotational lines probed here may be optically thick. The high optical depth is corroborated by predictions from numerical models \citep[e.g.,][]{bruderer2012} as well as by the low value of the observed $^{12}$CO/\thCO\ integrated intensity ratio (in cases where \thCO\ is detected: $\sim$3--10 for AB Aur and HD100546, see Table~\ref{t:spirelinefluxes}). Thus, the derived values for rotational temperature should not be interpreted as the kinetic temperatures of the gas: \Trot $\neq$ \Tgas. 

To compare our results with rotational diagrams from earlier work by \citet{meeus2013}, we derive rotational temperature values from linear regression fits to two separate energy ranges, chosen such that each of the two can be represented as well as possible by a straight line. The \Trot\ values of the warm components found in Fig.~\ref{fig:rotatdiag}a,b,c are consistent with those from fits to the PACS-only dataset from \citet{meeus2013}, i.e., the warmest of the two components for HD~100546 and the single-component fits for HD~97048 and AB~Aur. For HD~100546, the break between the cold and warm components is at the same energy of \Eup/$k$ = 1200~K both here and in \citet{meeus2013}. For the other two sources, the highest few lines in the SPIRE range fall on the same linear relation in the rotation diagram as that defined by the PACS measurements.

% suspicious CO 23-22 as upper limit
We have chosen not to include the CO 23--22 line (\Eup=1524~K) in this analysis because there are indications that this PACS detection may suffer from instrumental effects that lead to a systematic overestimate of the line flux. For example, \citet{meeus2013}  showed that CO 23--22 falls above the trend set by the other lines in all six sources of their sample of eight in which this line is detected, including AB~Aur and HD~97048. 

When the lines from the SPIRE~FTS measurements are added to the rotation diagrams (Fig.~\ref{fig:rotatdiag}a,b,c), it becomes apparent that additional, colder components need to be invoked for AB~Aur and HD~97048, at $\sim$80 and $\sim$20~K, respectively. In the case of HD~100546, for which \citet{meeus2013} already introduce a two-component fit, the colder of the two drops from $\sim$280 to 215~K by adding the SPIRE line measurements. The curvature of the trend plotted out by the observed data points suggests that even two discrete temperature components are insufficient for HD~100546. 

In conclusion, the SPIRE measurements of intermediate $J$ CO lines connect the ground-based lines (\Eup/$k$ $<$ 50~K) with the PACS lines  (\Eup/$k$ $>$ 500~K) in the rotation diagrams. A cooler component of CO gas is revealed that was not seen in the PACS-only study. As discussed above, most of the conditions on which the rotation diagram method is based do not hold for physical conditions prevailing in protoplanetary discs. In reality, molecular line transitions of different energies arise from (radially) different locations in the disc. More sophisticated models must be applied to take into account the detailed physical and chemical structure of discs (see Section~\ref{sec:prodimo}). 

Figure~\ref{fig:rotatdiag}d shows the rotation diagram of the T Tauri object DR~Tau, where two components are required to fit the SPIRE line detections. The spectrally resolved CO~3--2 measurement allowed \citet{thi2001b} to separate contributions from the disc and the extended envelope. The point shown in the rotation diagram represents purely the disc contribution. The adjacent, but spectrally unresolved, 4--3 line observed with SPIRE is also likely to suffer from contributions from the extended envelope, so we plot it as an upper limit in the rotation diagram. 
The derived rotational temperature of the warm component for DR~Tau is lower than for the Herbig Ae/Be objects in the other three panels. Confirmation of this difference should come from (non-)detections of higher $J$ CO lines in PACS observations (Blevins et al.,~in prep.), which probe the high energy range unaccessible to the ground-based and SPIRE observations presented here.

\subsection{Comparison with protoplanetary disc models}
\label{sec:prodimo}

\begin{figure}
    \includegraphics[width=0.5\textwidth]{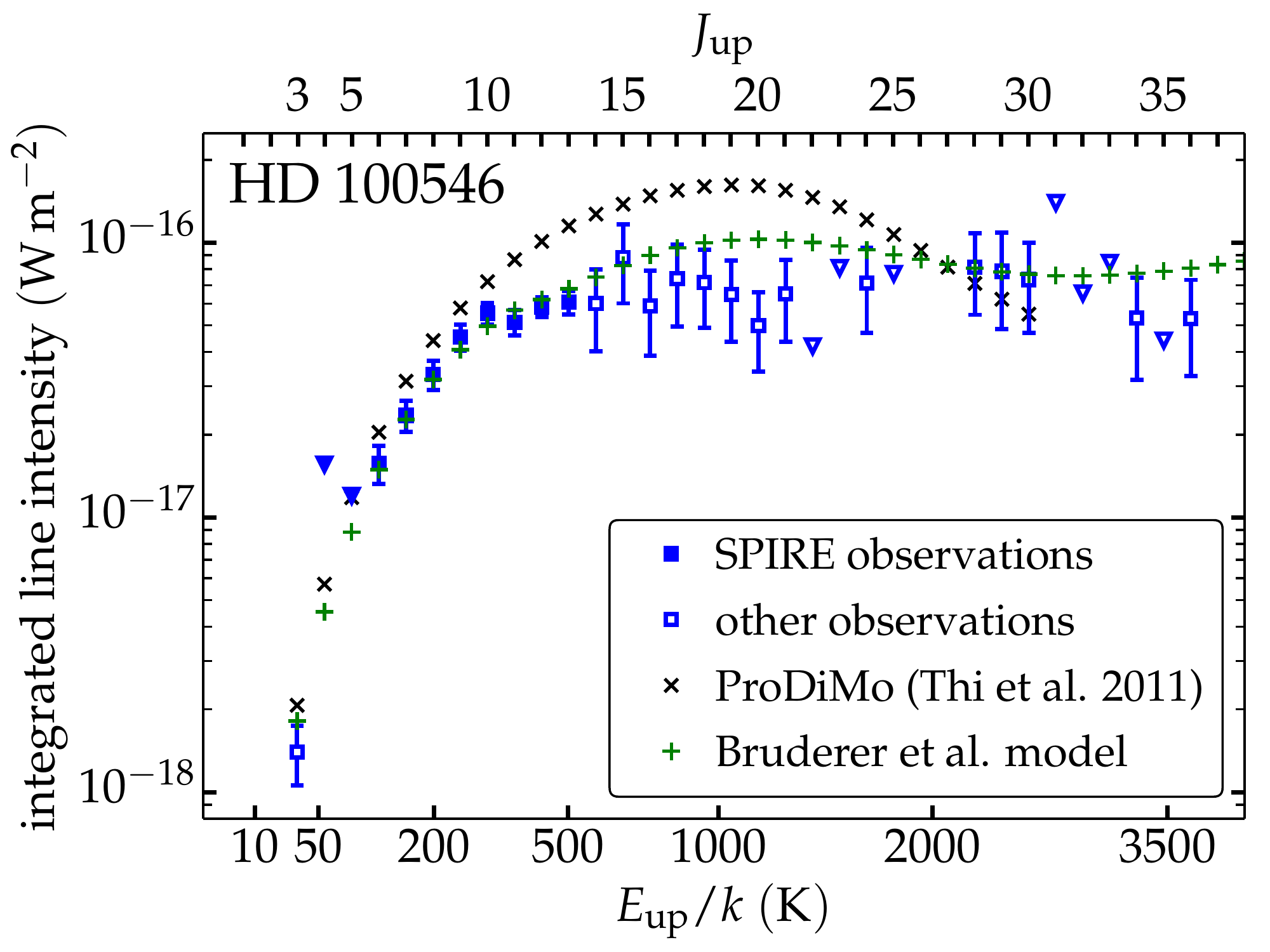} \\
    \includegraphics[width=0.5\textwidth]{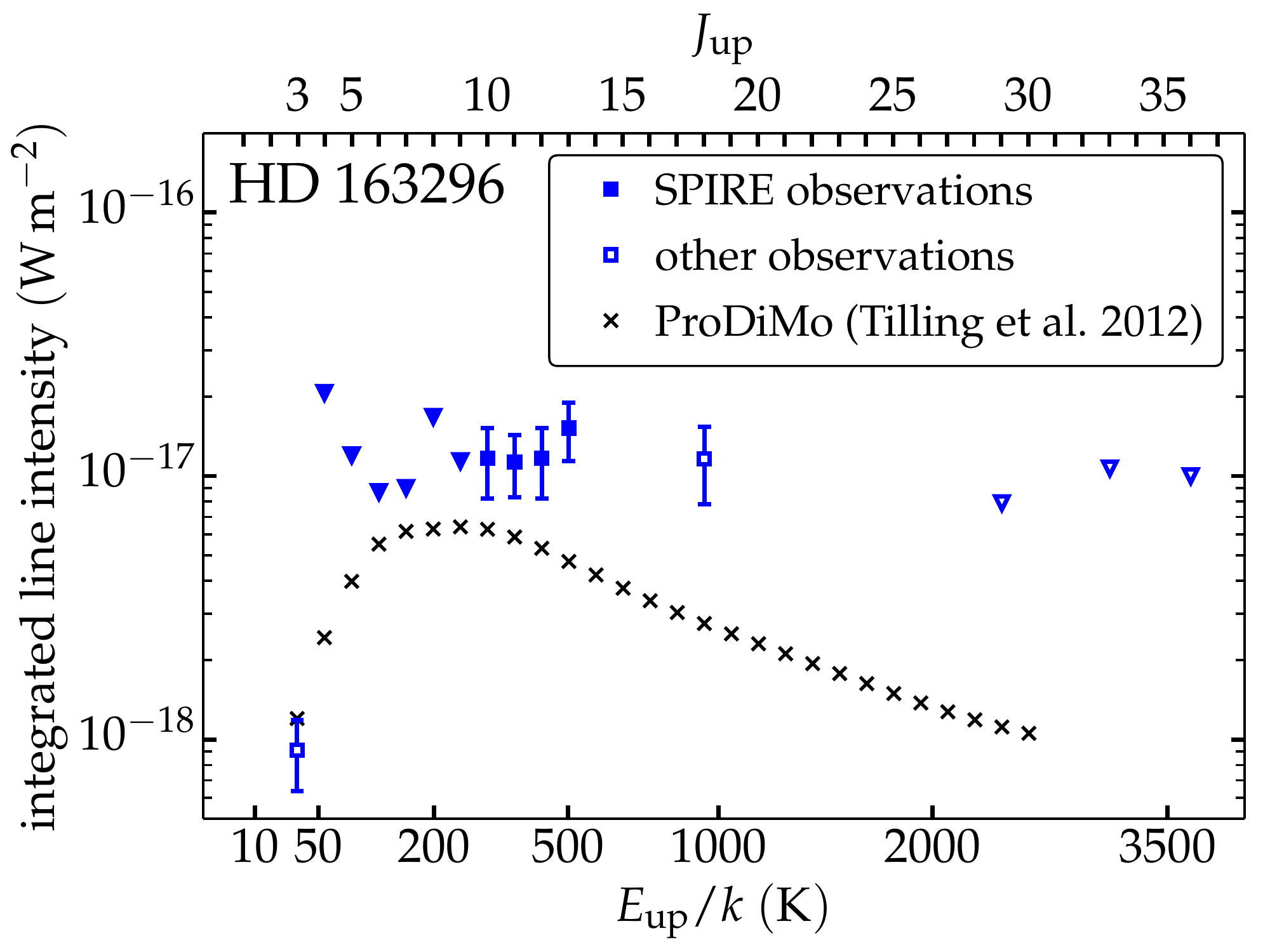}
    \caption{Observed CO spectral line energy distributions of the two targets with at least three CO lines detected in the SPIRE bands, for which a ProDiMo model is also available (black crosses). In addition, for HD~100546 we also show in green plus symbols the model provided by Bruderer (private communication), calculated directly from the model as published in \citet{fedele2013b}. Observed line detections are shown as (blue) squares, and 3-$\sigma$ upper limits are shown as triangles for lines that are either not detected or detected at $\leq$3~$\sigma$. Error bars represent the combination of line intensity errors (see Table~\ref{t:spirelinefluxes} for SPIRE) and the absolute calibration uncertainty, 6\% for SPIRE, 30\% for PACS \citep[see][section~4.10.2]{pacsobsmanual2013}. 
Data points with open markers are either ground-based (Appendix~\ref{sec:litdata}), i.e., at lower energies than the SPIRE data points, or from {\it Herschel} PACS measurements at higher energies (\citealt{meeus2013} for HD~100546;  \citealt{meeus2012} for HD~163296).
%The horizontal scale is linear in $\sqrt{E_\mathrm{up}}$ to obtain equal spacing between each consecutive rotational line transition. 
      }
    \label{fig:SLEDs}
\end{figure}

In this paper, we do not aim to provide an exhaustive explanation of the observations using complex protoplanetary disc models. Instead, we present in this section a comparison of the new SPIRE data of three sources for which detailed chemical-physical models have been published in the literature. These models of the discs around HD~100546, HD~163296, and TW~Hya are generated by the thermo-chemical protoplanetary disc model code ProDiMo (\citealt*{woitke2009a}; \citealt{kamp2010}) and are based on a large set of multiwavelength observational constraints. The spectral line energy distributions (SLED) for the first two of these targets are shown in Fig.~\ref{fig:SLEDs}. %For comparison, in Fig.~\ref{fig:SLEDs_nomodel} we also provide SLED diagrams for sources without models.
% CO 23-22 as upper limit
In any targets where the CO 23--22 is reported in PACS studies \citep{meeus2013}, we regard this value as an upper limit  (see Section~\ref{sec:rotatdiag}). 

\subsubsection{HD~100546}
\label{sec:prodimoHD100546}

\begin{figure}
    \includegraphics[width=0.5\textwidth]{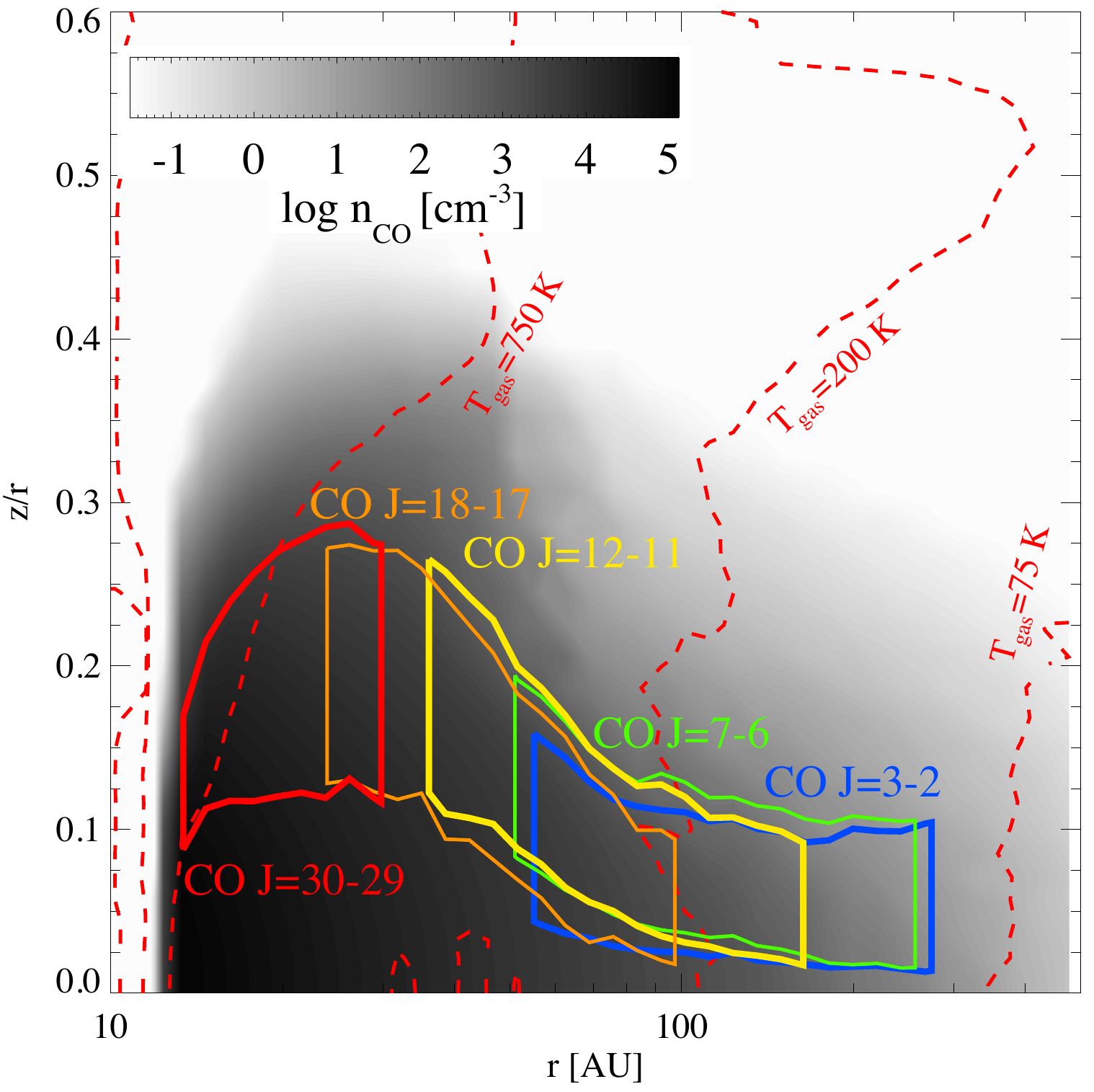}
    \caption{Illustration of the origin of several CO rotational lines in a model of the HD~100546 disc. Solid outlines represent areas delimited by the cumulative 15 and 85\% line flux contribution of each transition, both vertically and radially. The grey-scale indicates number density of CO molecules. Dashed contours mark gas temperature. The inner disc, at $<$10~au, is not shown, since it does not contribute significantly to the total amount of line flux from pure rotational transitions of CO. }
    \label{fig:lineorigin}
\end{figure}

The ProDiMo model for HD~100546 shown in the top panel of Fig.~\ref{fig:SLEDs} is drawn from work by \citet{thi2011}, and fits the near-infrared to millimetre continuum spectral energy distribution and near-infrared interferometric data of this source \citep{benisty2010b,tatulli2011}. The model reproduces the \Oi\ and \Cii\ fine structure lines, the \CHplus\ rotational lines \citep{thi2011}, low-$J$ CO emission, as well as the $v$=1--0 CO ro-vibrational lines in \citet{heinbertelsen2014}. 

% line origin figure
Figure~\ref{fig:lineorigin} shows the radially separated origin of the suite of rotational lines of CO from the model by \citet{thi2011}, illustrating that the CO rotation diagrams in Section~\ref{sec:rotatdiag} would not be expected to yield a single value for \Trot. The figure also confirms that the $^{12}$CO rotational lines are optically thick and originate in the upper layers of the outer disc.

Although the ProDiMo model for HD~100546 was not tailored to fit the intermediate- and high-$J$ range of the CO rotational ladder, the model points are always within a factor 3 of the observed CO line fluxes (Fig.~\ref{fig:SLEDs}, top panel). However, the observed SLED appears to level off around \Eup/$k$=200~K, roughly mid-way through the SPIRE range, whereas the current ProDiMo model continues to increase up to 1000~K. This may indicate that the (radial and vertical) temperature structure of the model disc needs to be adjusted. Specifically, since the CO rotational lines detectable by SPIRE and PACS are optically thick, the shape of the rotational SLED is sensitive to adjustments of the gas temperature in the relatively tenuous, irradiated upper layer of the disc, where PAHs are the dominant heating agent. 

Apart from the shape of the optically thick $^{12}$CO SLED, there is a larger discrepancy with respect to \thCO\ line strengths predicted by the ProDiMo model (not shown). Several \thCO\ lines between 6--5 and 11--10 are detected in HD~100546 (Table~\ref{t:spirelinefluxes}) at intensities of (6--17)$\times$$10^{-18}$ \Wsqm, while the \thCO\ lines predicted by the model are factors of $\sim$10--30 weaker. 
% increase 13CO by increasing disc gas mass
To test the dependence of modelled \thCO\ line intensities on total gas mass, we inspect the results from a model with a gas-to-dust ratio of 100, compared with $\sim$1 in the original model. We find that line intensities for \thCO\ increase by a factor 100, putting them well \emph{above} the measured \thCO\ rotational lines. On the other hand, the $^{12}$CO rotational line intensities increase by factors $\sim$10--30 ($<$100 due to higher optical depth) leading to a significant overestimate of the observations. An intermediate value for gas-to-dust ratio could bring the \thCO\ line intensities closer to the observed values, but would still grossly overestimate $^{12}$CO. 

% suggest isotope selective processes as partial explanation 
The inability of current models of HD~100546 to fit the $^{12}$CO and \thCO\ ladders simultaneously could be due to a combination of reasons. Besides the radial and vertical structure of \Tgas\ mentioned above, isotope selective photodissociation, which would suppress the \thCO\ abundance in low extinction regions \citep{visser2009b}, and chemical fractionation, which would affect the $^{12}$CO/\thCO\ ratio in colder layers deeper in the disc \citep{duley1984book}, could play a role. Neither of these isotope selective processes are implemented in the ProDiMo code. Future investigation is needed to determine which of the above processes is dominant in the \thCO\ line emitting regions, in order to explain the measured \thCO\ line strengths, expected to better trace total disc mass, while still retaining the good match of the more optically thick $^{12}$CO ladder in the current model.

For comparison, the top panel of Fig.~\ref{fig:SLEDs} also shows the CO SLED from a different model for the same source developed independently by \citet{bruderer2012} and most recently updated in \citet{fedele2013b}. Since these authors have focused attention on the CO ladder as observable with {\it Herschel} PACS and HIFI, their results \citep[from the model as presented in][]{fedele2013b} follow the shape of the observed SLED even closer than the ProDiMo description does, while also reproducing the broadband continuum spectral energy distribution \citep{mulders2011}. The approaches taken in each of the two codes are distinctly different in some aspects, for example with regard to unknown heating efficiency of small dust particles and PAHs, and adopted formation rates of \HH. In general, calculations by different PDR codes are prone to large differences in resultant gas temperature  \citep[see the comparison study by][]{rollig2007}. Small deviations in physical structure, radiative transfer, or chemistry are expected to translate non-linearly into changes in the gas temperature in the upper layers of the outer disc, and thus the resulting CO rotational line intensities. The gas-to-dust ratio adopted in the Bruderer et al.~model is 20 times higher than in the published ProDiMo model. 

Given our \thCO\ detections, detailed predictions for \thCO~line intensities using a time-dependent, isotopologue selective chemistry could reveal the true total gas mass in the disc of HD~100546. Such processes are currently being implemented within the framework of the model by \citeauthor{bruderer2012}, who have yet to publish predictions for \thCO\ line intensities. 
Further comparison of the different methods and assumptions used in the two independent codes may guide refinement of models, leading to predictions that better match all observational constraints.

\subsubsection{HD~163296}
\label{sec:prodimoHD163296}

The ProDiMo model points for HD~163296 shown in the bottom panel of Fig.~\ref{fig:SLEDs} are derived from the model defined by \citet{tilling2012}, which fits the optical to millimetre continuum spectral energy distribution, low-$J$ CO line intensities and spectral profiles from ground-based interferometric observations \citep{isella2007}. The CO SLED predicted by this model, however, falls a factor 2--4 short of the observed mid- to high-$J$ CO lines, including those presented in this work. The \water\ lines detected in the mid- and far-infrared with {\it Spitzer} IRS and {\it Herschel} PACS {\citep{pontoppidan2010a,fedele2012,meeus2012} are also underproduced by the ProDiMo model. This shortfall of specific molecular line intensities indicates that the gas at radii $\la$100~au is likely warmer than what is predicted in the \citet{tilling2012} ProDiMo model, as was also suggested by \citet{degregorio-monsalvo2013}.

\subsubsection{TW~Hya}
\label{sec:prodimoTWHya}

\citet{kamp2013} present a ProDiMo model for the disc surrounding the T~Tauri star TW~Hya. It reproduces the observed line strength of CO~18--17 presented in the same paper. In contrast, the 4-$\sigma$ detection of CO 13--12 for TW~Hya reported here (Table~\ref{t:upperlimits}) is almost 10 times higher than the intensity resulting from the \citeauthor{kamp2013} ProDiMo model. % and thus inconsistent to >3 sigma. 
This model's underestimation of the CO~13--12 line intensity for TW~Hya, however, should be interpreted with caution. The 13--12 line in the TW~Hya SPIRE SSW spectrum has a spectral signature consistent with the instrumental line shape, but the adjacent 12--11 and 11--10 lines are not detected and imply 3-$\sigma$ upper limits of roughly half the detected line intensity of $J$=13--12. These and other upper limits to CO line intensities from TW~Hya are not inconsistent with the ProDiMo model.

% -------- discussion ----------
\section{Discussion}
\label{sec:discussion}

\subsection{Mid-$J$ CO lines as a disc temperature probe}

In protoplanetary discs, the mid-$J$ CO lines observable by SPIRE (Tables~\ref{t:spirelinefluxes} and \ref{t:upperlimits}) are optically thick and originate from the disc's upper layers. Their intensities are therefore sensitive to the gas temperature in those regions (see Section~\ref{sec:prodimoHD100546}). We have divided our sample into T~Tauri objects and Herbig Ae/Be objects (see Table~\ref{t:sourceparam}), the latter with stronger stellar UV flux than the former. In addition, discs around Herbig Ae/Be stars are categorized depending on their flaring geometry \citep[e.g.,][]{meeus2001}. Compared to a flat structure, such a flaring geometry would lead the disc surface to intercept considerably more UV flux and thereby heat up more the disc's upper layers. In contrast, X-rays irradiation, which is typically stronger for T~Tauri discs than for Herbig Ae/Be discs, seems to have little effect on modelled CO chemistry and thermal balance (\citealt{aresu2012}; Aresu, private communication). Therefore, assuming a UV driven thermal balance, flaring Herbig discs are expected to show the strongest mid-$J$ CO emission. In this subsection, the CO $J$=10--9 line is taken as representative for the mid-$J$ CO ladder and is compared with other observational metrics to derive qualitative conclusions. 

First, it appears that in our sample the flaring Herbig discs are generally brightest in both mid-$J$ CO and the local continuum (near 250~\micron), whereas most T~Tauri discs and flat (non-flaring) Herbig discs are less bright. This conclusion is tentative, however, given that our sample consists of only four to eight objects in each of the three categories. 
Secondly, \mbox{CO~10--9} may be correlated with both CO 18--17 and  \Oi\ $^3$P$_1$--$^3$P$_2$ line emission (at 63\micron, observed with PACS, \citealt{meeus2012,meeus2013,kamp2013,fedele2013a,howard2013,thi2010a}). With only four or five targets in our sample detected in these lines, however, there are too few data points to be conclusive. A statistical correlation between CO 10--9 and \Oi\ 63\micron\ would be interesting, given that these two lines are seen to originate from roughly the same spatial region in ProDiMo models.

\subsection{HCN $J$=11--10 detections}

Compared to $^{12}$CO, the hydrogen cyanide molecule (HCN) emitting area lies deeper in a protoplanetary disc, since it only becomes abundant in regions where the gas density and UV extinction are both high \citep[e.g.,][]{jonkheid2007} and its rotational transitions have higher critical densities than those of CO. We report here detections of the \mbox{$J$=11--10} transition of HCN at 974 GHz towards three of the flaring Herbig Ae/Be discs: HD~169142, AB~Aur and HD~100546, though only at $<$~3~$\sigma$ (Tables~\ref{t:spirelinefluxes} and \ref{t:upperlimits}). In HD~169142, the ratio beween the integrated intensities of the brightest CO line and HCN~11--10 is $\sim$1, whereas the same ratio in AB~Aur and HD~100546 is between 4 and $\sim$20. 

To our knowledge, no ground-based (sub)millimetre HCN detections have been reported in the literature for HD~169142 and HD~100546, so the $J$=11--10 SPIRE measurements constitute the first detections of a rotational line of HCN towards these two objects. For AB~Aur, \citet{fuente2010b} report a detection of HCN~3--2 with the EMIR receiver at the IRAM~30-m Telescope, with a spectral line profile suggestive of a disc origin. Their integrated line intensity converts to \pow{4}{-21}~\Wsqm, yielding an \mbox{HCN~11--10/3--2} ratio of $\sim$500--2500. 
Finally, the SPIRE spectrum of TW~Hya yields a 3-$\sigma$ upper limit of \pow{8}{-18}~\Wsqm\ for the integrated intensity of the HCN~11--10 line, but the HCN 4--3 line \emph{was} detected towards the same target \citep[\pow{1.1}{-19} \Wsqm,][]{vanzadelhoff2001}. The resulting \mbox{HCN~11--10/4--3} ratio is constrained at $\la$ 100. 

Future work employing detailed physical-chemical models is warranted to investigate what information the HCN rotational lines provide on the denser, deeper parts of protoplanetary discs. More sensitive observational constraints at various submillimetre frequencies, for example from ALMA, could help to construct a full HCN SLED, much like the ones shown for $^{12}$CO in this work that probe the tenuous upper disc layers. Contrary to CO, the HCN rotational lines are split into hyperfine components and, specifically in the lower $J$ states of HCN, care should be taken to account for the effects of hyperfine line anomalies \citep{loughnane2012}.

\subsection{\CHplus\ in HD~100546 and HD~97048}
\label{sec:discussCHplus}

The \CHplus\ $J$=1--0 line at 835 GHz is not detected in any of the 18 targets studied in this paper. Upper limits for its intensity are listed in Tables \ref{t:spirelinefluxes} and \ref{t:upperlimits}. Here we highlight HD~100546 and HD~97048, for which the 3-$\sigma$ upper limits for the lowest rotational transition of \CHplus\ are \pow{6}{-18} and \pow{1.6}{-17} \Wsqm, respectively. Both of these values are consistent with the trends defined by the complementary higher energy \CHplus\ line detections and upper limits (\Jup=2 to 6) found by \citet{thi2011} and \citet{fedele2013a} using the PACS spectrometer. The new upper limit for \mbox{$J$=1--0} thus cannot resolve the question whether \CHplus\ emission originates from the inner rim of the outer disc of HD~100546, as suggested by the ProDiMo approach in \citet{thi2011}, or from further out in the disc, as indicated by the slab model by \citet{fedele2013a}.

\subsection{Upper limits on low-excitation \water\ lines}
\label{sec:discusswater}

The 3-$\sigma$ limits to the integrated intensities of low-excitation \water\ lines between 500 and 1500~GHz (Table~\ref{t:spirelinefluxes}), including the ground-state transitions of both ortho- and para-\water, are $\sim$$10^{-17}$~\Wsqm\ for most targets, and $\sim$\pow{6}{-18}~\Wsqm\ for targets with longer exposures such as TW~Hya,  DR~Tau, FZ~Tau, VW~Cha and RNO~90. For the sources with existing ProDiMo models (HD~100546 (Section~\ref{sec:prodimoHD100546}), HD~163296 (Section~\ref{sec:prodimoHD163296}) and TW~Hya (Section~\ref{sec:prodimoTWHya})), predicted \water\ line intensities are consistent with the measured upper limits in Table~\ref{t:spirelinefluxes}. In addition, the {\it Herschel} HIFI detections of p-\water\ 1$_{11}$--0$_{00}$ (\pow{6}{-19}~\Wsqm) and o-\water\ 1$_{10}$--1$_{01}$ (\pow{2}{-19}~\Wsqm) in TW~Hya \citep{hogerheijde2011} are both comfortably below the upper limits derived from our SPIRE spectrum.

\subsection{Neutral carbon emission towards HD~100546}
\label{sec:carbonHD100546}

Table~\ref{t:spirelinefluxes} lists a \Ci\ $^3$P$_2$--$^3$P$_1$ integrated intensity at 809~GHz (370~\micron) of (7.6$\pm$3)~$10^{-18}$~\Wsqm\ towards HD~100546. This detection is inconsistent with the upper limit of $<$\pow{0.85}{-18}~\Wsqm\ obtained from ground-based observations with the APEX observatory by \citet{panic2010}, but our SPIRE measurement is believed to be superior (Section~\ref{sec:spire-heterodyne}). If this \Ci\ emission originates from the disc, the comparatively low C/O ratio employed in \citet{bruderer2012}, \citet{bruderer2013} and \citet{fedele2013b}, inspired by the APEX upper limit, may no longer be necessary. In fact, the SPIRE \Ci\ intensity is roughly consistent with the \pow{4}{-18}~\Wsqm\ in the `representative model' of \citet{bruderer2012}. On the other hand, the ProDiMo model for HD~100546 \citep[see also Section~\ref{sec:prodimoHD100546}]{thi2011,heinbertelsen2014} predicts a \Ci\ 809~GHz line flux of \pow{17}{-18}, considerably closer to the SPIRE line detection than to the earlier non-detection by \citet{panic2010}.  

One possibility for reconciling the two apparently inconsistent observations of the \Ci\ $^3$P$_2$--$^3$P$_1$ line is that the SPIRE measurement may include a contribution from a component other than the disc of HD~100546.  
This explanation, however, is deemed unlikely for the following two reasons. First, a fully diffuse component would likely extend over several arcminutes on the sky and would therefore have been caught by the off-centre detectors in the SPIRE array (see Section~\ref{sec:linedetections}), which it was not. Secondly, a relatively compact envelope type structure (within $\sim$30\arcsec) that may have escaped the SPIRE off-centre detectors is unlikely to emit significantly in \Ci\ lines, as already argued by \citet{bruderer2012}. % (their sect. 5.3)
The best way to resolve the apparent inconsistency between the two observations is to obtain a more sensitive, spectrally resolved observation of the line that could also identify the kinematic origin of any detected line signal.

\section{Conclusions}
\label{sec:conclusions}

This paper presents {\it Herschel} SPIRE spectroscopic data in the continuous 450--1540~GHz range (666--195~\micron) of a sample of 18 protoplanetary discs. Of these spectra, six targets show a significant amount of detectable spectral line signal, while most other targets only exhibit continuum emission. The spectral line detections are dominated by ten consecutive rotational lines of mid-$J$ $^{12}$CO ($J$=4--3 up to 13--12), and also include low signal-to-noise signatures from \thCO, \Ci\ and HCN (Section~\ref{sec:lineresults}). The CO transitions observed with SPIRE trace rotational energy levels between $\sim$50 and 500~K. Augmented with observations of lower and higher energy transitions from the literature, the CO rotational ladder is compared with existing, published physical-chemical models of discs, where available (Section~\ref{sec:analysis}). 

From the collected data and models, we find the following. 
\begin{enumerate}
	\item The $^{12}$CO spectral line energy distribution (SLED) of the disc around Herbig Ae/Be star HD~100546 is optically thick across the SPIRE frequency range and is well matched, within a factor 3, by two completely independent model codes. 
	\item The \thCO\ line detections in our SPIRE spectrum of HD~100546 exceed the values predicted by the ProDiMo model by factors $>$10. It is not straightforward to adjust the model to scale to the \thCO\ observations without compromising the match to the \twCO\ counterpart. Isotope selective (photo)chemical processes may play a role, and should be investigated in more detail in numerical models.
	\item In the sample, composed of 12 Herbig Ae/Be objects and 6 T~Tauri objects, the brightest mid-$J$ CO emission is typically observed in discs around Herbig objects, specifically those with flaring disc geometries. In contrast, there are also Herbig objects with little to no detectable line signal in our SPIRE observations. In addition, two of the six T~Tauri discs do show significant CO line emission, although these detections are aided by the deeper integrations towards these targets. 
	\item There may be a correlation between the cooling power through the \Oi~63~\micron\ line and that of the CO rotational lines, but our collection of overlapping data points is too sparse to be definitive. 
	\item Besides the \twCO\ and \thCO\ lines, we tabulate for all 18 targets upper limits at the frequencies of \CHplus\ \mbox{($J$=1--0)} and of eight low-energy \water\ transitions, to be used as constraints for future physical-chemical modelling. 
	\item The SPIRE observation of HD~100546 reveals a detection of \Ci~$^3$P$_2$--$^3$P$_1$ at 809 GHz (370~\micron) that is inconsistent with a previous ground-based measurement with the APEX observatory \citep{panic2010}. The {\it Herschel} observation presented in this work (see Table~\ref{t:spirelinefluxes} and Sections~\ref{sec:spire-heterodyne} and \ref{sec:carbonHD100546}) does not suffer from atmospheric attenuation and has better accuracy than the earlier ground-based one.
\end{enumerate}

\section*{Acknowledgements}

% CSA and NSERC
MHDvdW and DAN are supported by the Canadian Space Agency (CSA) and the Natural Sciences and Engineering Research Council of Canada (NSERC). 
% EU FP7 for Inga, Wing-Fai and Peter
IK, WFT and PW acknowledge funding from the European Union Seventh Framework Programme FP7-2011 under grant agreement no.~284405. 
% SPIRE
SPIRE has been developed by a consortium of institutes led by Cardiff University (UK) and including Univ.~Lethbridge (Canada); NAOC (China); CEA, LAM (France); IFSI, Univ. Padua (Italy); IAC (Spain); Stockholm Observatory (Sweden); Imperial College London, RAL, UCL-MSSL, UKATC, Univ.~Sussex (UK); and Caltech, JPL, NHSC, Univ. Colorado (USA). This development has been supported by national funding agencies: CSA (Canada); NAOC (China); CEA, CNES, CNRS (France); ASI (Italy); MCINN (Spain); SNSB (Sweden); STFC, UKSA (UK); and NASA (USA).
% Gibion, Hugh, Ros, Simon, Giambattista 
We acknowledge Gibion Makiwa and Hugh Ramp for early data processing efforts and initial exploration of the GT1 dataset, and Rosalind Hopwood for running the adjusted pipeline to process the five observations with low CEV temperatures. We thank Simon Bruderer for providing his model predictions for HD~100546 and for discussions on gas physics in discs, and Giambattista Aresu for discussions on UV and X-ray induced chemistry in discs. 
% referee
We warmly acknowledge the thoughtful review provided by the anonymous referee.
% NASA ADS:
This research has made use of NASA's Astrophysics Data System Bibliographic Services. 
% astropy:
This research made use of Astropy, a community-developed core Python package for Astronomy (Astropy Collaboration \citeyear{astropy2013}), and the matplotlib plotting library \citep{matplotlib2007}.

\bibliographystyle{mn2e_latex/mn2e.bst}
% ! use version of mn2e.bst by Michael Williams (2010)
%    not the one distributed with MNRAS package (1995)
%    latter does not deal with papers of >8 authors4
\footnotesize
\bibliography{../../literature/allreferences}
\normalsize

% APPENDIX 
\appendix

% non-SPIRE data
\section{Auxiliary data from the literature}
\label{sec:litdata}

This paper makes use of auxiliary spectral line data of the targets in our sample observed with ground-based facilities.
Table~\ref{t:grounddata} lists CO line intensities of these (sub)millimetre transitions gathered from the literature. Since `radio' intensities observed with heterodyne instruments are commonly expressed in brightness temperature units, K for intensity or \Kkms\ for integrated intensity, we convert the values listed in the source papers to \Wsqm. The conversion, suitable for pointlike sources, uses the Rayleigh-Jeans law to convert K to $\mathrm{W}~\mathrm{m}^{-2}~\mathrm{Hz}^{-1}$ \citep{baars1973} and $\mathrm{d}\nu/\nu = \mathrm{d}v/c$ to convert \kms\ to Hz: 
\begin{equation}
\label{eq:KkmstoWm2}
\int I_\nu \mathrm{d}\nu \ [\mathrm{W\,m}^{-2}] = \frac{2 k}{A_\mathrm{geom} \eta_\mathrm{a}} \frac{\nu}{c} \int {T_A}^\star \, \mathrm{d}v ,
\end{equation}
where $k$ is the Boltzmann constant in J~K$^{-1}$, $A_\mathrm{geom}$ is the geometric collecting area of the telescope in m$^2$ ($\pi (D/2)^2$), \etaa\ is the telescope's aperture efficiency (and $A_\mathrm{eff} = \eta_\mathrm{a} A_\mathrm{geom}$ the effective telescope aperture; see \citealt*{wilson2009book}), $\nu$ is the frequency of the observed spectral line in Hz, and $c$ is the speed of light in \kms. The integrated line intensity, $\int {T_A}^\star \, \mathrm{d}v$ (in \Kkms), is deliberately defined in the ${T_A}^\star$ scale, which for a point source is the quantity that couples to the source intensity \citep[e.g.,][]{baars1973}. In many cases, the literature reports intensities in \Tmb\ (`main beam' brightness temperature scale), which is converted to \Tastar\ = (\etamb/$\eta_\mathrm{l}$)~\Tmb\ before substituting into Eq.~\ref{eq:KkmstoWm2}. For JCMT, we take $\eta_l = 1$, and for APEX $\eta_l = 0.97$ \citep{guesten2006}. Values for \etaa\ and \etamb\ are given in the notes to Table~\ref{t:grounddata}.

\begin{table}
\caption{Ground-based (sub)millimetre measurements of CO lines from the literature (see notes below the table). Where the literature values are listed in \Kkms\ or Jy~\kms\ units, the values are converted as described in the text of Appendix~\ref{sec:litdata}. In addition to the uncertainties listed in square brackets in this Table, absolute calibration uncertainties are added for the various observatories: 10\% for \mbox{IRAM~30-m}, 30\% for JCMT RxB3 and 20\% for APEX-2a. The exception is the CO 3--2 line intensity of DR~Tau, for which \citet{thi2001b} already included calibration uncertainties in their tabulated uncertainty values. 
}
%(4): \citet{panic2008} and \citet{raman2006}. Source value already in Jy scale: 12CO 2--1: 12.1 Jy \kms, from SMA, emission summed over central 4$\times$4\arcsec\ area. \\
%(5): \citet{oberg2011a}, SMA: source value already in Jy scale: 12CO 2--1 is 20.76 [0.23] Jy~\kms. \\
%(6):  \citet{degregorio-monsalvo2013}, ALMA: source value already in Jy scale. \\
%(7): \citet{qi2011}, SMA: source value already in Jy scale. \\
\label{t:grounddata}
\begin{tabular}{l l r l}
\hline
Object		& Transition 	& $\int I_\nu \, \mathrm{d}\nu$ & Reference \\
			&			& ($10^{-18}$ \Wsqm) &  \\
\hline
AB Aur 		& CO 2--1 & 0.9 [0.4]	& ($a$) \\
			& CO 3--2 & 1.56 [0.04]	&  ($b$) \\  %  JCMT CO 3-2 flux detected by Thi+ (2001) is *much* larger than that published by Dent+ 2005; and Lin+ (2006, SMA) only detect 20% of the Thi et al. value 
HD 100546	& CO 3--2 & 1.4 [0.2] 	& ($c$) \\
HD 163296	& CO 3--2 & 0.91 [0.02] 	& ($b$)   \\
%			& CO 2--1 & 0.397 [0.003] & (7) \\
%			&               & 1.157 [0.001] & (6) \\
%			& CO 6--5  & 1.353 [0.149] & (7) \\ 
%HD 169142 	& CO 2--1 & 0.089 [0.003] & (4) \\
%			& CO 3--2 & 0.36 [0.03] 	& ($b$) \\
%HD 142527 	& CO 2--1 & 0.152 [0.002]	& (5)  \\
DR~Tau		& CO 3--2 & 1.1 [0.3] & ($d$) \\
\hline
\end{tabular} 
\begin{minipage}{0.5\textwidth}
References~--- 
($a$) \citet{tang2012}, IRAM~30-m Telescope. Source value already in Jy. Adopting conservative error of $\sim$50\%, since value was gauged from a printed figure. 
($b$) \citet{dent2005}, JCMT RxB3: $D=15.0$~m; \etamb=0.62; \etaa=0.53. 
($c$) \citet{panic2010}, APEX-2a: $D=12.0$~m; \etamb=0.73; \etaa=0.60. 
($d$) \citet{thi2001b}, JCMT RxB3: $D=15.0$~m; \etamb=0.62; \etaa=0.53. Only the two spectral components that are ascribed to the DR~Tau disc by \citet{thi2001b}. 
\end{minipage}
\end{table}

\label{lastpage}
\end{document}